\documentclass[twocolumn]{aastex62}

\usepackage{subfigure,epsfig,amsfonts}
\usepackage{natbib}
\usepackage{amsmath}
\usepackage{amssymb}
\usepackage{amsthm}
\usepackage{graphicx}
\usepackage{commath}

\newcommand{\Kepler}{\emph{Kepler\ }}
\newcommand{\Mjup}{M_\mathrm{Jupiter}}

\usepackage{comment}

\def\mymathhyphen{{\hbox{-}}}

\begin{document}

\correspondingauthor{Sean M. Mills}
\email{sean.martin.mills@gmail.com}

\author{Sean M. Mills}
\affiliation{California Institute of Technology, Department of Astronomy\\
1200 East California Blvd, Pasadena, CA 91125, USA \\
}

\author{Andrew W. Howard}
\affiliation{California Institute of Technology, Department of Astronomy\\
1200 East California Blvd, Pasadena, CA 91125, USA \\
}

\author{Lauren M. Weiss}
\affiliation{Institute for Astronomy, University of Hawaii \\
2680 Woodlawn Drive, Honolulu, HI 96822, USA\\
}
\affiliation{Parrent Fellow}

\author{Jason H. Steffen}
\affiliation{University of Nevada, Las Vegas, Department of Physics and Astronomy\\
4505 S Maryland Pkwy, Las Vegas, NV 89155\\
}

\author{Howard Isaacson}
\affiliation{Department of Astronomy, University of California\\
510 Campbell Hall, Berkeley, CA 94720, USA\\
}

\author{Benjamin J. Fulton}
\affiliation{California Institute of Technology, Department of Astronomy\\
1200 East California Blvd, Pasadena, CA 91125, USA \\
}
\affiliation{IPAC-NASA Exoplanet Science Institute \\
Pasadena, CA 91125, USA \\
}

\author{Erik A. Petigura}
\affiliation{California Institute of Technology, Department of Astronomy\\
1200 East California Blvd, Pasadena, CA 91125, USA \\
}

\author{Molly R. Kosiarek}
\affiliation{Department of Astronomy and Astrophysics, University of California\\
Santa Cruz, CA 95064, USA\\
}
\affiliation{NSF Graduate Research Fellow}

\author{Lea A. Hirsch}
\affiliation{Department of Astronomy, University of California\\
510 Campbell Hall, Berkeley, CA 94720, USA\\
}
\affiliation{Kavli Institute for Particle Astrophysics and Cosmology\\
Stanford University, Stanford, CA 94305, USA\\
}

\author{John H. Boisvert}
\affiliation{University of Nevada, Las Vegas, Department of Physics and Astronomy\\
4505 S Maryland Pkwy, Las Vegas, NV 89155\\
}

\title{Long-Period Giant Companions to Three Compact, Multiplanet Systems}
\shorttitle{Compact Multis with Giant Companions}

\begin{abstract}
Understanding the relationship between long-period giant planets and multiple smaller short-period planets is critical for formulating a complete picture of planet formation. This work characterizes three such systems. We present Kepler-65, a system with an eccentric ($e=0.28\pm0.07$) giant planet companion discovered via radial velocities (RVs) exterior to a compact, multiply-transiting system of sub-Neptune planets. We also use precision RVs to improve mass and radius constraints on two other systems with similar architectures, Kepler-25 and Kepler-68. In Kepler-68 we propose a second exterior giant planet candidate. Finally, we consider the implications of these systems for planet formation models, particularly that the moderate eccentricity in Kepler-65's exterior giant planet did not disrupt its inner system.
\end{abstract}

\section{Introduction}
\label{sec:intro}

Precise photometric monitoring of more than 100,000 stars over a four year period by the \Kepler space mission \citep{2010Sci...327..977B} has revealed thousands of planets and planetary candidates \citep{2016ApJS..224...12C,2016ApJ...822...86M}. Approximately half of these planets and candidates are found in systems of two or more transiting planets with orbital periods $\lesssim$1 year. However, the existence of long-period or inclined planetary companions is rarely probed by the \Kepler data alone \citep[c.f.,][]{2012Sci...336.1133N,2014ApJ...791...89D}. Therefore, we rely on radial velocity (RV) follow-up to probe the full architectures of planetary systems observed by \emph{Kepler}. Here we present results on three systems that have multiple transiting planets discovered by \Kepler as well as long-period giant planet companions detected via RVs as part of the California Planet Search \citep[CPS;][]{2010ApJ...721.1467H}: Kepler-25, Kepler-65, and Kepler-68. 

\begin{figure}
\centerline{
\includegraphics[scale=0.6]{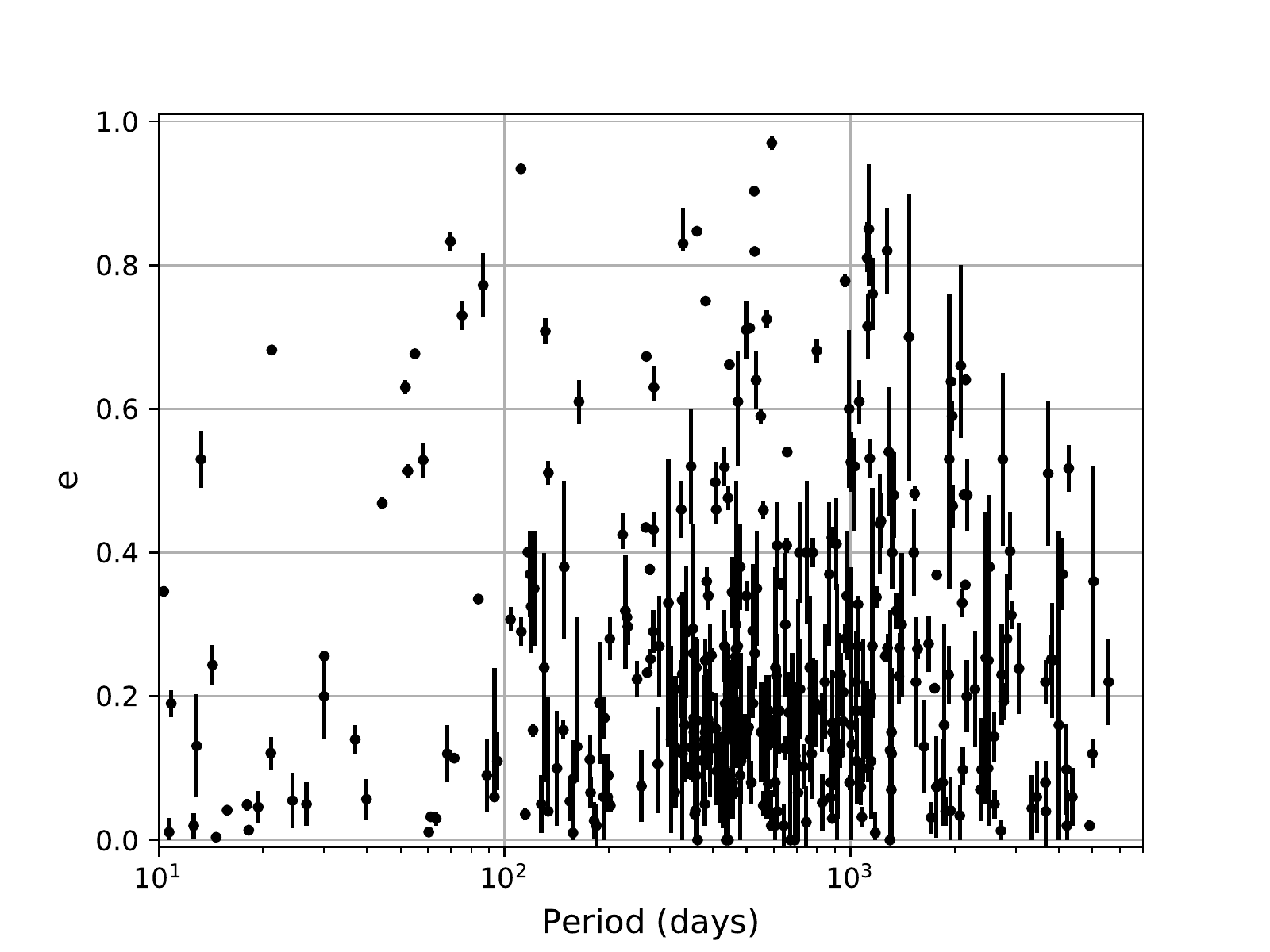}
}
\centerline{
\includegraphics[scale=0.6]{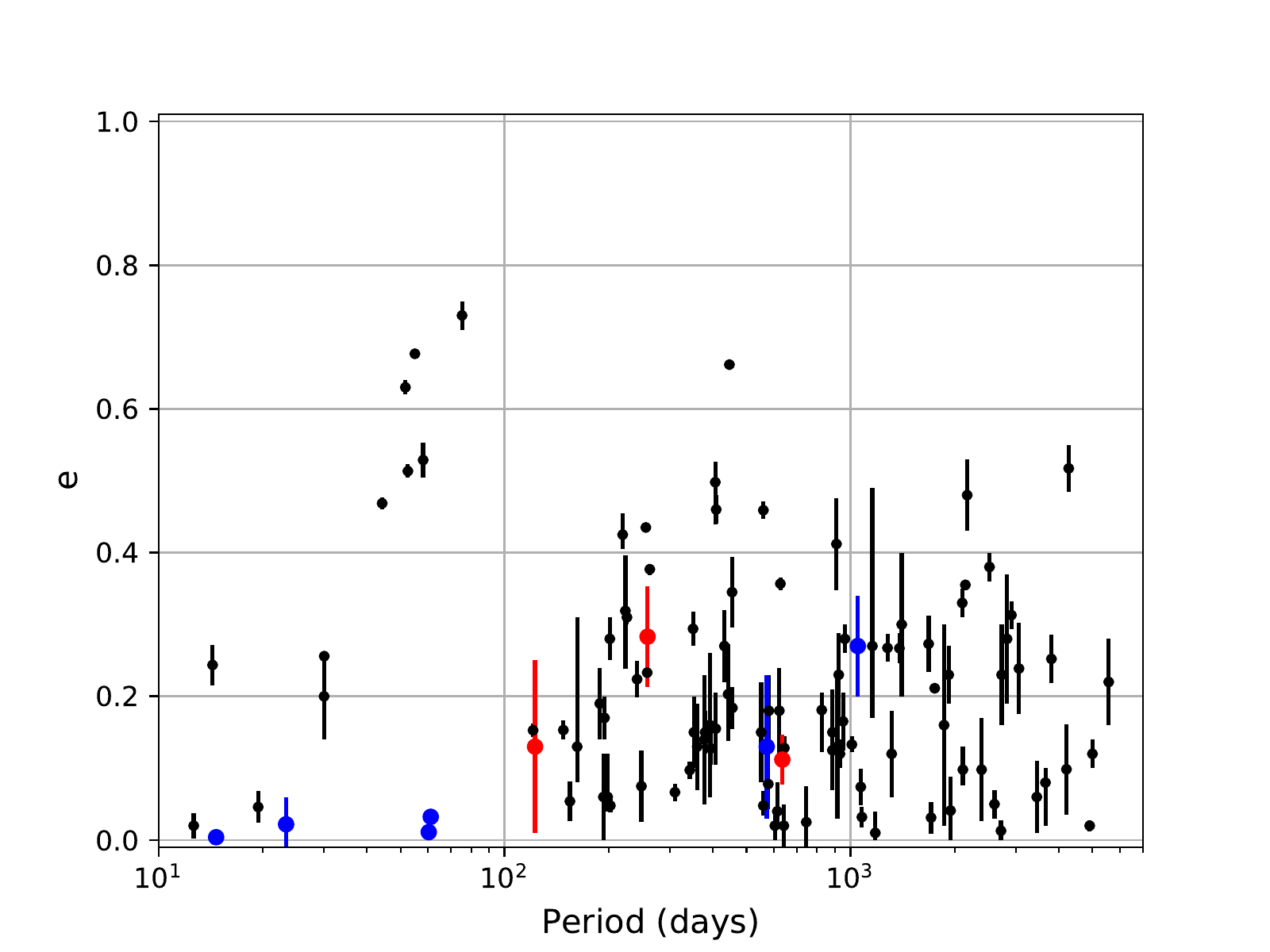}
}
\caption{\emph{Top:} The periods and eccentricities of warm and cool Jupiters: planets with $M \sin i>0.5$ $\Mjup$ and $P>10$ days that have eccentricity and eccentricity uncertainty data in the Exoplanet Orbit Database (http://exoplanets.org) as of October 2018 \citep{2014PASP..126..827H}. \emph{Bottom:}  The eccentricities of giant planets in multi-planet systems with measured eccentricities are shown in black. The majority are systems of planets with exactly two giant planets. Blue points highlight those planets with multiple small ($M < M_{Saturn}$ or $R < 8 R_\oplus$) planets in the same system. The red points are a subset of those points which are the subject of this study. \label{fig:wcje}
}
\end{figure}

How giant planets exterior to super-Earths and sub-Neptunes (SEASNs) affect their formation and evolution is an open question in planetary science. It is speculated that Jupiter and Saturn prevented the typical short-period SEASNs observed in Kepler multiplanet systems from forming in the Solar System \citep{2015PNAS..112.4214B}, with similar effects possible in exoplanet systems \citep{2015ApJ...800L..22I}. This effect remains uncertain and in general it seems that the presence of warm and cool Jupiters does not prevent such compact SEASN architectures, and may even favor them \citep{2018AJ....156...92Z,2018arXiv180608799B}. Counterexamples include Kepler-90, a pair of coplanar, $e\approx0$ giant planets exterior to 5 SEASNs \citep{2014ApJ...781...18C} and WASP-47, a system with a $e\approx 0.27$, $\sim$600 day Jupiter around a mixed system of 2 SEASNs and an $e\approx0$ Jupiter at $<$10 days -- although the latter example may be atypical due to its relatively short-period Jupiter in addition to its exterior one \citep{2015ApJ...812L..18B,2017AJ....153...70S}.

Warm and cool Jupiters (WCJs; $P\gtrsim 10$ days and $M\gtrsim0.5$ $\Mjup$) have a very broad eccentricity distribution (Fig.~\ref{fig:wcje}). The mean eccentricities ($\bar{e}$) of these Jupiters is $\approx0.3$, much higher than the Solar System planets ($\bar{e}\approx0.06$) and Kepler multi-planet systems as measured via transit duration ratios \citep[Rayleigh $\sigma_e \approx 0.03$;][]{2014ApJ...790..146F}, using asteroseismology derived stellar densities \citep[$\sigma_e \approx 0.05$;][]{2015ApJ...808..126V}, combining transit durations and spectroscopy \citep[$\bar{e} \approx 0.04$;][]{2016PNAS..11311431X}, and via dynamical analyses \citep[$\bar{e} \approx 0.02$;][]{2017AJ....154....5H}. However, the WCJ $e$ distribution is perhaps similar to systems with a single transiting warm super-Earth or sub-Neptune \citep[$\bar{e} \sim 0.3$;][]{2016PNAS..11311431X,2019AJ....157...61V}.

While data are presently limited, it generally seems that systems with both multiple SEASNs and WCJs have dynamically cool Jupiters (e.g., Kepler-46, Kepler-148, Kepler-30, Kepler-89, Kepler-487, 55 Cnc, GJ 876, HD 34445, WASP-47, and Kepler-90)\footnote{This list is incomplete as it only includes well-characterized Jupiters reported on exoplanets.org. For instance, Kepler-48, a long-period Jupiter around 3 SEASNs from \citet{2014ApJS..210...20M}, is not included.}. Fig.~\ref{fig:wcje} compares the distribution of measured giant planet eccentricities in different types of multiple planet systems. The low eccentricities of giants in planets with multiple SEASNs is intriguing compared to the general population of WCJs, both in single and multiple planet systems, where high eccentricities are common. \citet{2016ApJ...822....2U} and \citet{2016AJ....152..206F} also tentatively suggests that \Kepler SEASN systems have coplanar (and thus dynamically cold) WCJ companions by showing evidence of potential $>$2-year period Jupiter companions\footnote{We note  \citet{2016ApJ...822....2U} also presents a few candidates of either eccentric long-period Jupiters or possibly Warm Jupiters based on transit duration, but this is uncertain.}. Even if WCJs are not barriers to planet formation, eccentric exterior Jupiters may excite any interior planets' eccentricities and inclinations causing a reduction in either true planet multiplicity from ejections or observed multiplicity due to increased planet mutual inclinations \citep{2017AJ....153..210H,2018MNRAS.478..197P}.

\section{Methods}
\label{sec:methods}

In order to characterize the Kepler-25, Kepler-65, and Kepler-68 systems, we rely on a combination of the transit photometry, RVs, and precise stellar spectra. The photometric signals reveal the ratio of planetary to host star radii. With precise stellar properties derived from combining stellar spectra and parallax measurements \citep{2018AJ....156..264F}, the physical sizes of the planets can be determined. Additionally the planet-planet gravitational interactions of closely spaced planets results in transit timing variations (TTVs) which reveal a combination of the mass and eccentricity of the planets \citep{2005MNRAS.359..567A,2012ApJ...761..122L}. Modeling these TTVs can give mass constraints on the planets; however, with low signal-to-noise (S/N) data a mass-eccentricity degeneracy often remains \citep{2012ApJ...761..122L,2015ApJ...802..116D}. These degeneracies can sometimes be broken by RV data that provide a complementary constraint \citep[e.g.,][]{2018AJ....156...89P}, or by a strong prior on eccentricity \citep{2017AJ....154....5H}. We study Kepler-65 and Kepler-25 with a photodynamical model which produces synthetic RV and photometric data generated by the N-body interactions of simulated planets and compares it to the observed data to extract planetary orbital elements, masses, and radii self-consistently. Kepler-68 does not exhibit detectable TTVs because the planets are not closely-spaced or near resonance \citep{2016ApJ...818..177A,2016ApJS..225....9H,2018ApJS..234....9O}, so we model only the RVs for that system.

\begin{figure}
\centerline{
\includegraphics[scale=0.5]{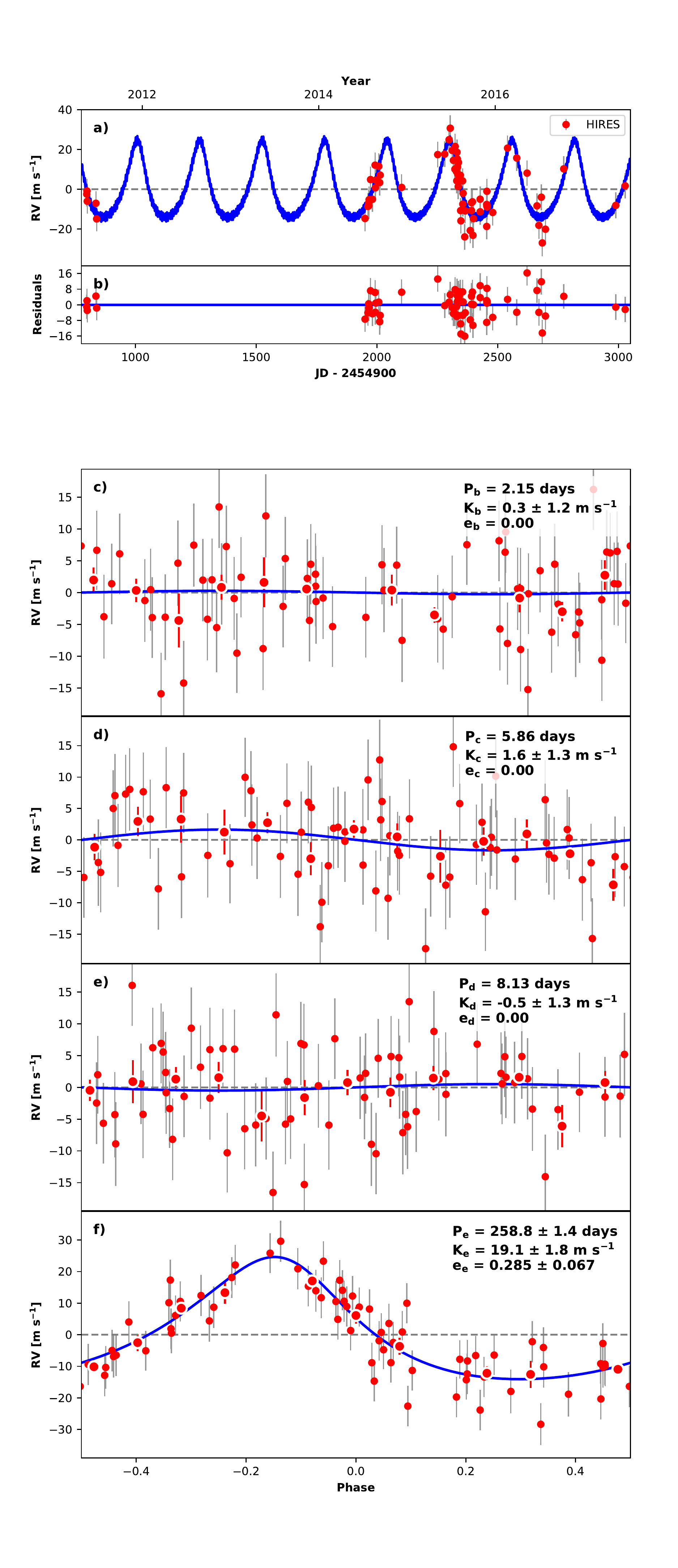}
}
\caption{\emph{Top Panel:} Kepler-65 RV \texttt{radvel} best-fit model and residuals. \emph{Lower Panels:} RV data for each planet phase-folded at the best-fit orbital period with all other planet's signal removed.\label{fig:k65rvbest}
}
\end{figure}

\begin{deluxetable}{lrr}
\tablecaption{Kepler-65 MCMC Posteriors\label{table:k65}}
\tablehead{\colhead{Parameter$^a$} & \colhead{Median$^{\mathrm{84}^{th}\mathrm{ Percentile}}_{\mathrm{16}^{th}\mathrm{ Percentile}}$} & \colhead{Unit}}
{\def\baselinestretch{0.95}
\startdata
\hline
\multicolumn{3}{c}{\bf Fitted Parameters}\\
\hline
P$_b$              & 2.1549209$^{+8.6e\mymathhyphen06}_{-7.4e\mymathhyphen06}$  &    days   \\
T$_0,$$_b$              & 801.32382$^{+0.00011}_{-0.00011}$  &  BJD-2454900     \\
$\sqrt{e}\cos \omega$$_b$       & -0.03$^{+0.16}_{-0.15}$  &       \\
$\sqrt{e}\sin \omega$$_b$       & -0.02$^{+0.12}_{-0.12}$  &       \\
i$_b$                   & 92.2$^{+1.3}_{-1.4}$  &    degrees   \\
M$_b$          & 0.0076$^{+0.0077}_{-0.0051}$  &   $\Mjup$    \\
R$_b$/R$_\star$        & 0.009215$^{+6.2e\mymathhyphen05}_{-4.8e\mymathhyphen05}$  &       \\
P$_c$              & 5.859697$^{+9.3e\mymathhyphen05}_{-9.9e\mymathhyphen05}$  &   days    \\
T$_0,$$_c$              & 803.39284$^{+0.00019}_{-0.00019}$  &    BJD-2454900    \\
$\sqrt{e}\cos \omega$$_c$       & 0.065$^{+0.073}_{-0.112}$  &       \\
$\sqrt{e}\sin \omega$$_c$       & 0.082$^{+0.079}_{-0.09}$  &       \\
i$_c$                   & 92.33$^{+0.29}_{-0.26}$  &   degrees     \\
M$_c$          & 0.017$^{+0.0054}_{-0.0052}$  &    $\Mjup$    \\
R$_c$/R$_\star$         & 0.016743$^{+7.8e\mymathhyphen05}_{-7.1e\mymathhyphen05}$  &       \\
P$_d$              & 8.13167$^{+0.00024}_{-0.00021}$  &    days   \\
T$_0,$$_d$              & 802.80285$^{+0.00042}_{-0.00037}$  &   BJD-2454900     \\
$\sqrt{e}\cos \omega$$_d$       & -0.05$^{+0.11}_{-0.07}$  &       \\
$\sqrt{e}\sin \omega$$_d$       & 0.01$^{+0.1}_{-0.08}$  &       \\
i$_d$                   & 92.35$^{+0.18}_{-0.16}$  &    degrees    \\
$\Omega$$_d$            & -0.13$^{+0.76}_{-0.71}$  &   degrees    \\
M$_d$          & 0.013$^{+0.0025}_{-0.0025}$  &  $\Mjup$      \\
P$_e$              & 258.8$^{+1.5}_{-1.3}$  &    days   \\
T$_0,$$_e$              & 1045.4$^{+6.8}_{-8.2}$  &    BJD-2454900    \\
$\sqrt{e}\cos \omega$$_e$       & 0.492$^{+0.067}_{-0.085}$  &       \\
$\sqrt{e}\sin \omega$$_e$       & 0.16$^{+0.12}_{-0.13}$  &       \\
i$_e$                   & 127.0$^{+27.0}_{-25.0}$  &    degrees    \\
$\Omega$$_e$            & -10.0$^{+130.0}_{-110.0}$  &       \\
M$_e$          & 0.82$^{+0.63}_{-0.16}$  &   $\Mjup$     \\
M$_\star$                   & 1.248$^{+0.018}_{-0.021}$  &   $M_{\odot}$     \\
R$_\star$                   & 1.437$^{+0.032}_{-0.027}$  &  $R_{\odot}$          \\
c$_1$                   & 0.349$^{+0.045}_{-0.047}$  &       \\
c$_2$                   & 0.227$^{+0.067}_{-0.064}$  &       \\
$\sigma$$_\mathrm{jitter}$            & 6.05$^{+0.75}_{-0.67}$  &   m s$^{-1}$    \\
$\gamma$ &   $0.92^{+0.93}_{-0.92}$     &  m s$^{-1}$ \\
\hline
\multicolumn{3}{c}{\bf Derived Parameters}          \\
\hline
$M_{jup,e} \sin i_e$    & 0.653$^{+0.056}_{-0.055}$  &   $\Mjup$     \\
M$_b$                   & 2.4$^{+2.4}_{-1.6}$  &     $M_\oplus$   \\
M$_c$                   & 5.4$^{+1.7}_{-1.7}$  &     $M_\oplus$   \\
M$_d$                   & 4.14$^{+0.79}_{-0.8}$  &    $M_\oplus$    \\
M$_e$                   & 260.0$^{+200.0}_{-50.0}$  &      $M_\oplus$  \\
R$_b$                   & 1.444$^{+0.037}_{-0.031}$  &    $R_\oplus$    \\
R$_c$                   & 2.623$^{+0.066}_{-0.056}$  &    $R_\oplus$    \\
R$_d$                   & 1.587$^{+0.040}_{-0.035}$  &    $R_\oplus$    \\
$\rho$$_b$              & 4.4$^{+4.5}_{-3.0}$  &   g cm$^{-3}$    \\
$\rho$$_c$              & 1.64$^{+0.53}_{-0.51}$  &   g cm$^{-3}$     \\
$\rho$$_d$              & 5.7$^{+1.2}_{-1.2}$  &    g cm$^{-3}$    \\
$e_b$                   & 0.028$^{+0.031}_{-0.02}$  &       \\
$e_c$                   & 0.02$^{+0.022}_{-0.013}$  &       \\
$e_d$                   & 0.014$^{+0.016}_{-0.010}$  &       \\
$e_e$                   & 0.283$^{+0.064}_{-0.071}$  &       \\
\enddata
}
\tablenotetext{a}{Osculating orbital elements are valid at $T_\mathrm{epoch} = 2,455,715$ BJD.
}
\end{deluxetable}

\subsection{HIRES Observations}

The radial velocity data for this work were collected using the HIRES spectrometer at the Keck Observatory from April 2010 to September 2017. The setup used for the RV observations was the same as used by the CPS \citep{2010ApJ...721.1467H}, with a resolving power of R $\approx$ 60,000 between wavelengths 3600 and 8000 \AA\ \citep{2008PhST..130a4001M,2014ApJS..210...20M}.

The Doppler analysis is the same as that used by the CPS group \citep{2010PASP..122..701J}. Template spectra obtained without the iodine cell were used to forward model the spectra taken with the iodine cell. The wavelength scale, the instrumental profile, and the RV in each of $\sim$700 segments of length 80 pixels in length corresponding to $\sim$2.0 \AA\ (depending on position along each spectral order) are all solved for simultaneously. The internal uncertainty in the final RV measurement for each exposure is the weighted uncertainty in the mean RV of those segments, with weights inversely proportional to the relative RV scatter of each segment \citep{2014ApJS..210...20M}. 

To measure and remove contaminating light from the sky, we used the C2 decker on HIRES which has an 0$''$.87 $\times$ 14$''$.0 field of view on the sky. The C2 decker simultaneously collects both the stellar light and night-sky light, and the sky-contamination is recorded along with the stellar spectrum at each wavelength in the regions above and below each spectral order. These ``sky pixels'' provide a direct measurement of the sky spectrum at that wavelength and are subtracted from the stellar spectrum to mitigate sky contamination.

\subsection{Photometric Modeling}

The \Kepler photometry is prepared for dynamical fits by detrending the simple aperture photometry (SAP) flux data from the \Kepler portal on the Mikulski Archive for Space Telescopes (MAST). For long-cadence data, we fit the amplitudes of the first five cotrending basis vectors away from transits to determine a baseline. We discard points whose quality flag had a value greater than or equal to 16. For short-cadence data, cotrending basis vectors are not available. For both cadences, we masked out the expected transit times plus 20\% of the full duration of each transit to account for possible timing variations. We then fit a cubic polynomial model with a 1-day width centered within half an hour of each data point to determine its baseline. We divide the flux and uncertainties by this baseline. A small amount of correlated noise was still present in the data likely due to known spurious instrumental frequencies \citep{2011MNRAS.414L...6G,keplerdatahandbook} and stellar variability. To avoid distorting the transit shapes, we do not attempt to detrend this short time-scale noise . We multiplied the normalized uncertainties from the \Kepler data set by the $\sqrt{ \chi^2_{nom} / N}$, where $N$ is the number of data points in the \Kepler data, and $\chi^2_{nom}$ is the $\chi^2$ of a nominal best-fit model. Thus the reduced $\chi^2$ of our best-fitting models was 1.00. By increasing our uncertainties, we conservatively widen our posteriors to take into account the scatter introduced by the aforementioned unmodeled noise in the system.

\subsection{Fits Only Using Radial Velocities}
\label{sec:rvmethods}

For the radial velocity fits, we use \texttt{radvel} \citep{2018PASP..130d4504F}, a Keplerian multi-planet radial velocity fitting routine. It is coupled to \texttt{emcee} \citep{2013PASP..125..306F}, a Markov Chain Monte Carlo (MCMC) engine, so that Bayesian posteriors may be found. Each RV fit includes a radial velocity jitter term, $\sigma_{\mathrm{jitter}}$ to take into account stellar and instrumental noise above the measured uncertainty level \citep[for details, see][]{2018PASP..130d4504F}. The planets's orbits were fitted in a basis of \{$K$, $P$, $T_0$, $\sqrt{e} \cos \omega$, $\sqrt{e} \sin \omega$\}, where $K$ is the radial velocity amplitude, $P$ is the planetary period, $T_0$ is the time of planet-star conjunction, and $e$ and $\omega$ are planetary eccentricity and argument of pericenter respectively. We use uniform priors on each parameter unless explicitly stated. For each model we run an MCMC to fit the data and determine uncertainties, stopping once all parameters have a Gelman-Rubin statistics of $<1.01$ \citep{Gelman92}. We convert the measured RV amplitudes into absolute planet masses by applying stellar mass constraints from \citet{2018AJ....156..264F}.

\subsection{RV-TTV Fits}
\label{sec:ttvmethods}
We use a photodynamic model to produce theoretical lightcurves for comparison to the \Kepler photometry. The model takes initial conditions at a specified epoch, and integrates the N-body equations of motions of the planets over the interval of the data. If the integration reaches time of an RV data point, the stellar RV is computed by summing the effects of all planets at that instant. Every time a planet passes in front of the star, a theoretical lightcurve is computed at each \Kepler time-series data point using the model described in \citet{2012MNRAS.420.1630P}. For long-cadence data, we computed the flux value at 15 equally spaced points in time over a cadence's integration and averaged them together to produce the model value. The model and data are then compared in a $\chi^2$ sense to compute the likelihood of the model, accounting for any priors for the parameters (specified for each system below). We perform a differential evolution (DE) MCMC \citep{TerBraak2005} in order to understand the uncertainties of the system parameters. 

The parameters for each planet in our model are the osculating Jacobian orbital elements $P$, $T_0$, $\sqrt{e} \cos \omega$, $\sqrt{e} \sin \omega$, $i$, $\Omega$, $M/M_\star$, and $R/R_\star$, where $\Omega$ is the longitude of the ascending node, and $M$ and $R$ are mass and radius respectively. We model the star with four free parameters: $R_\star$, $M_\star$, and two quadratic limb-darkening parameters $c_1$ and $c_2$, following \citet{2012MNRAS.420.1630P}. $\sigma_\mathrm{jitter}$ is a final free parameter.

\begin{figure}
\centerline{
\includegraphics[scale=0.45]{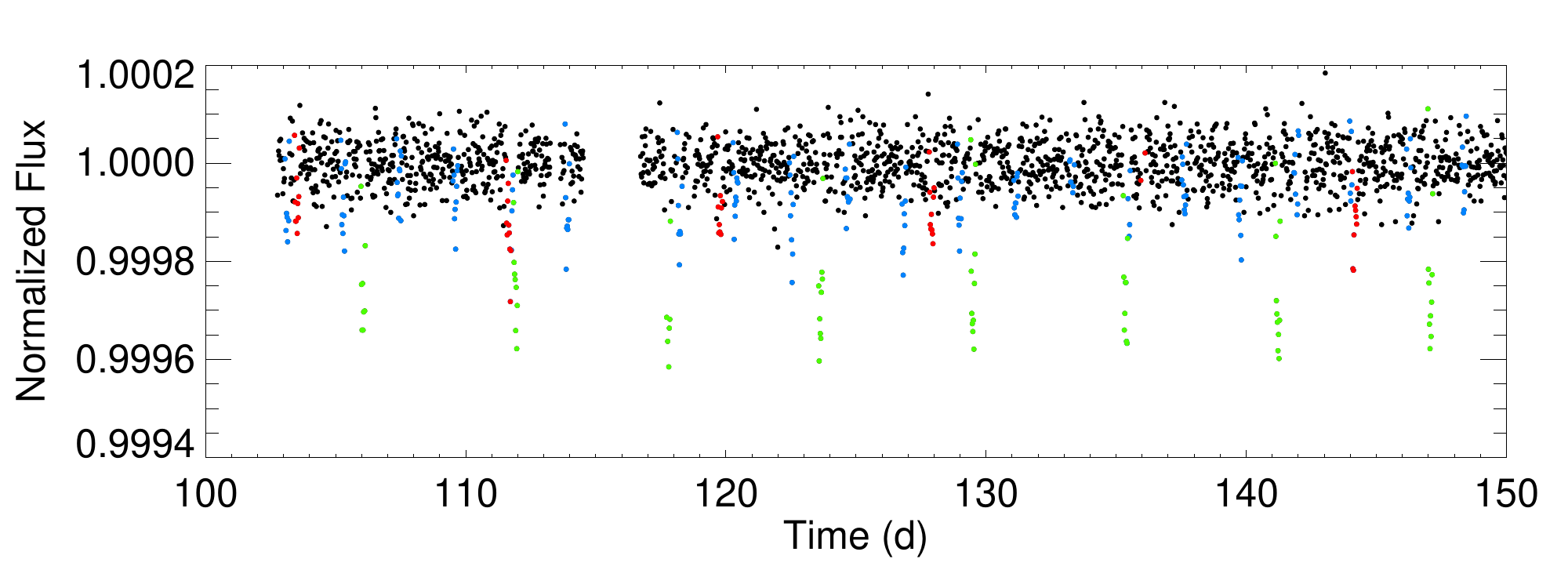}
}
\caption{A segment of the long-cadence portion of Kepler-65's detrended lightcurve (BJD-2454900) with transit events highlighted in blue (planet b), green (planet c), and red (planet d). \label{fig:k65lc}
}
\end{figure}

\begin{figure}
\centerline{
\includegraphics[scale=0.5]{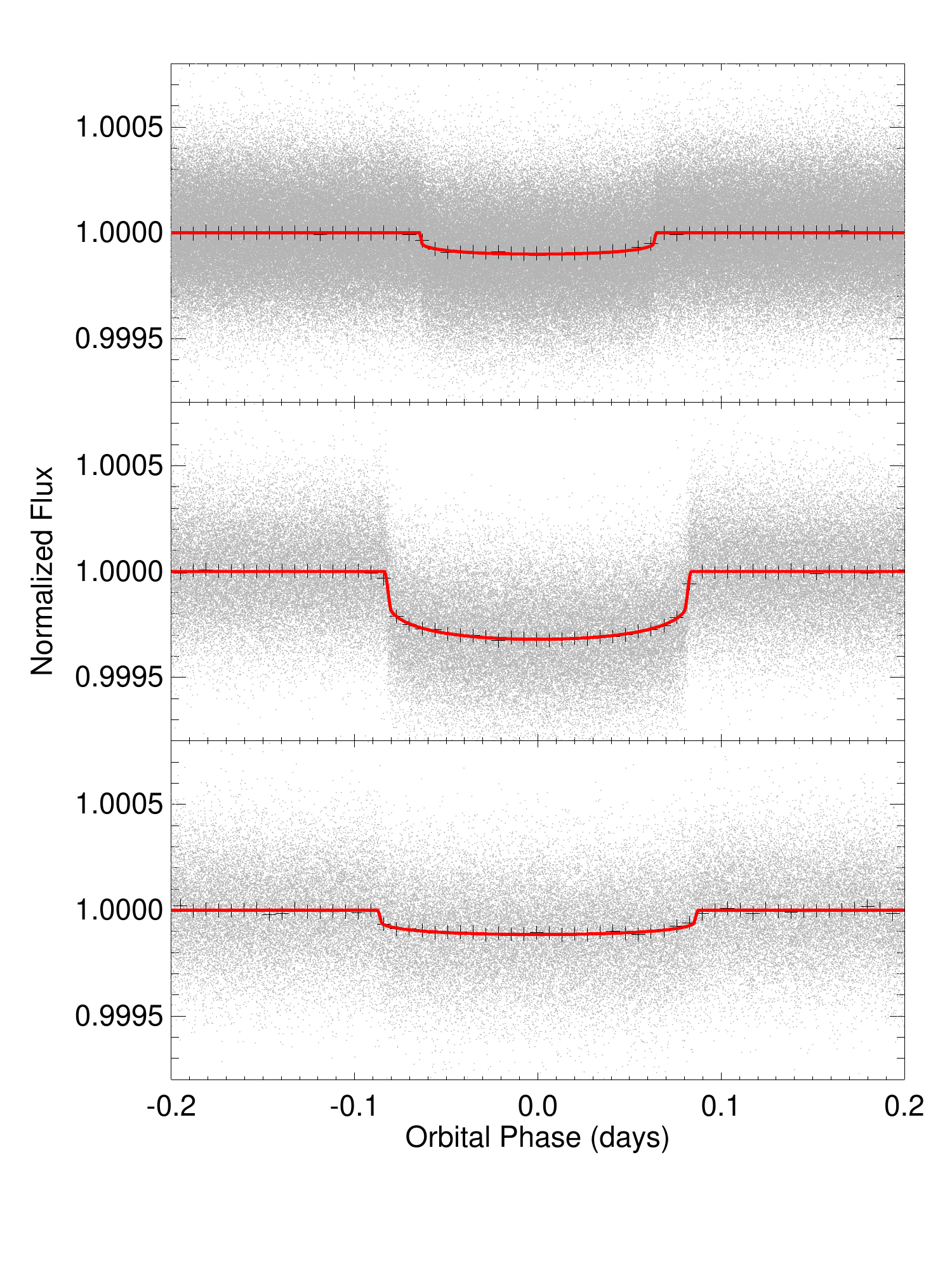}
}
\caption{ Kepler-65's three transiting planets b, c, and d are show from top to bottom phase-folded with TTVs removed. Gray points are individual \Kepler data, black crosses are data binned in 10 minute intervals, and the red line is a best fit transit model. \label{fig:k65tranfold}
}
\end{figure}

\begin{figure}
\centerline{
\includegraphics[scale=0.6]{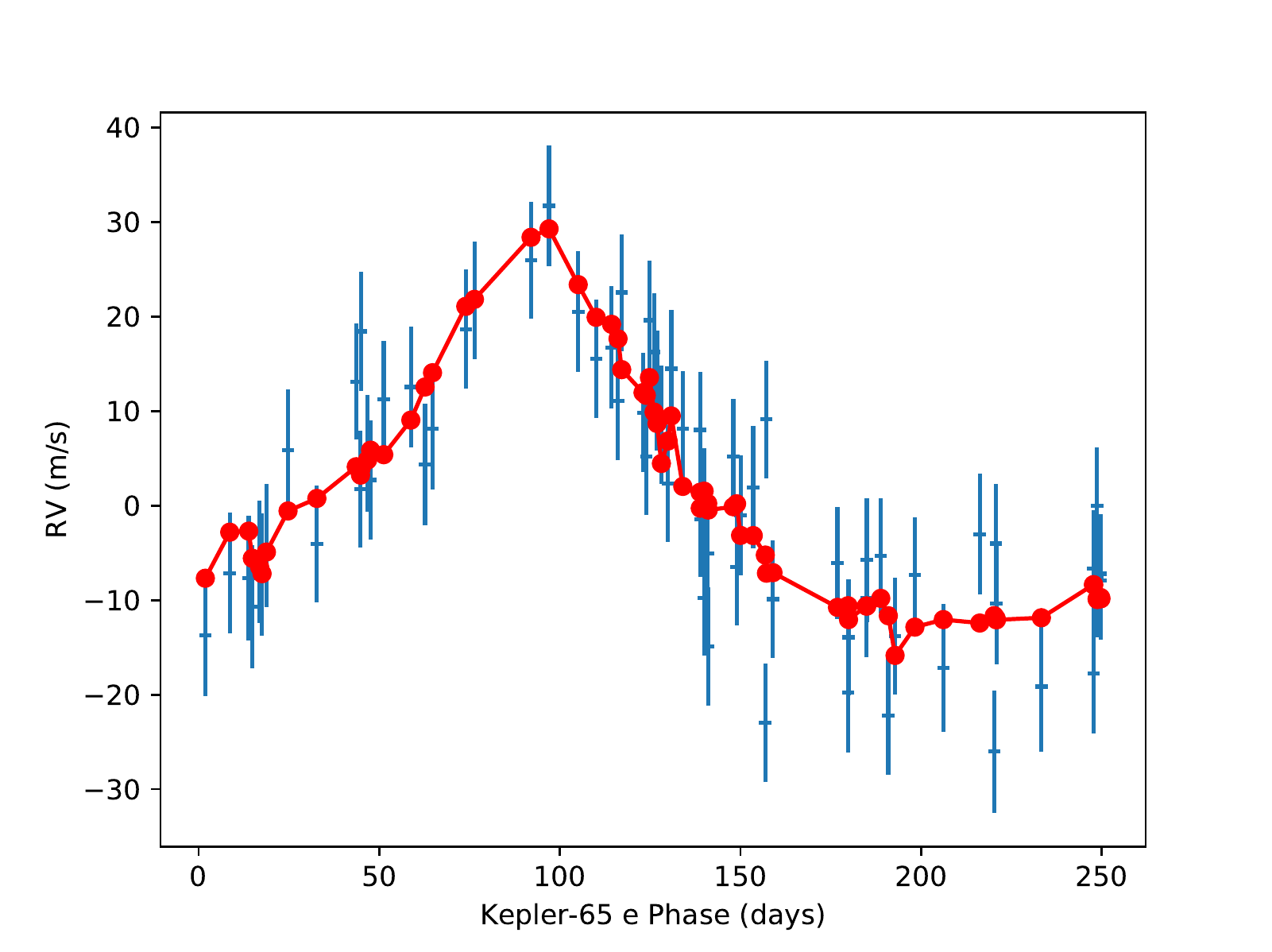}
}
\centerline{
}
\caption{ 
Kepler-65's best RV fit (red) and the HIRES RV data (crosses with uncertainties including the best-fit RV jitter of 5.3 m s$^{-1}$) phased at the best-fit orbital period of the long-period giant planet e. The small amplitude variations in the theoretical points is due to the RV contribution of the 3 inner planets, whose RV signatures are no longer completely periodic in an N-body model. Their low amplitudes agree with the RV-only fit. The deviation of the large-amplitude shape from a sine curve belies the eccentricity in the giant planet. 
\label{fig:k65emass}
}
\end{figure}

\begin{figure}
\centerline{
\includegraphics[scale=0.6]{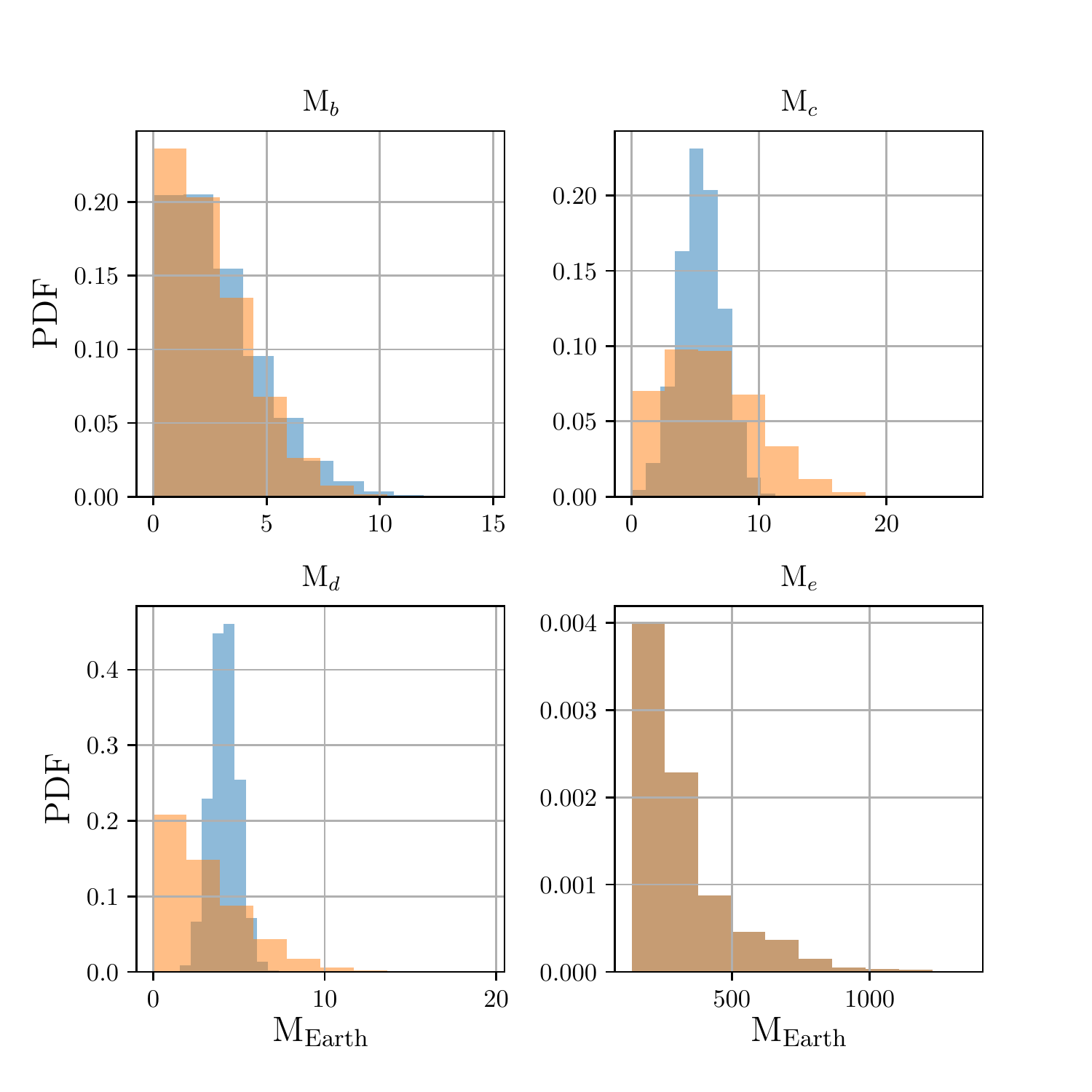}
}
\caption{Mass posteriors for all planets in the Kepler-65 system from the joint RV-\Kepler photometry fit (blue) and from RVs alone (orange). While the inner planet (b) has an upper limit from RVs alone, the addition of the transit timing variations significantly improves the precision of planet c and d's masses. \label{fig:k65masses}
}
\end{figure}

\section{Kepler-65}

Kepler-65 (KOI 85, KIC 5866724) is a system of three transiting planets all with $R<3 R_\oplus$ and orbital periods $<10$ days as originally validated by \citet{2013ApJ...766..101C}. \citet{2013ApJ...766..101C} showed that the inner planets' orbits are aligned with the spin of the star. \citet{2014ApJ...787...80H} marginally detected a TTV signal which implies a non-zero mass for planet c using the first 12 quarters of long-cadence data \citep{2013ApJS..208...16M}. HIRES RV data revealed an additional companion which we denote with the subscript $e$ at an orbital period of $258.7^{+1.5}_{-1.3}$ days and $M \sin i$ of $0.67 \pm 0.06$ $\Mjup$. Below we explore a fit to the radial velocities alone and to the photometry and RVs simultaneously for higher precision determination of the system's architecture.


\subsection{Kepler-65 RV Only Fit}

We perform a Keplerian RV fit with \texttt{radvel} \citep{2018PASP..130d4504F} on the HIRES RV data set (Table~\ref{table:k65rvs}). The inner planets have well determined orbital periods from \emph{Kepler}. Additionally, they are likely to have low eccentricities \citep{2015ApJ...808..126V}, so in this preliminary RV-only fit we fix their eccentricities to 0 to prevent degeneracies with mass. Therefore the only free parameters for the inner 3 planets are the $K$ amplitudes (i.e., their masses). We allow all parameters of the newly discovered outer planet to vary ($P$, $T_0$, $\sqrt{e}\sin\omega$ , $\sqrt{e}\cos\omega$, $K$). Our best fit model is shown in Fig.~\ref{fig:k65rvbest}. None of the inner planets are detected at high significance.  However, the non-transiting planet is detected at $>10\sigma$ significance, and the departure from a perfectly sinusoidal shape of the outer planet's RV signal reveals a significant eccentricity ($e=0.286\pm0.069$). We note that removing the giant planet's eccentricity in the RV fit results in an increase in the Akaike Information Critera by 12.1, which is sufficient to strongly rule out the $e=0$ model \citep{1974ITAC...19..716A,burnham2003model}. 

We note that a Lomb-Scargle (LS) periodogram \citep{1976Ap&SS..39..447L,1982ApJ...263..835S} of the data indicates a unique, well defined peak at 259 days, eliminating the possibility the observed periodicity is an alias for a different period of the giant planet induced by the time sampling. We also consider stellar activity as a potential false positive for an apparent planetary signal. A periodogram of the S-values of the HIRES RV data does not show a peak at the period of the putative planet and there is almost no correlation ($\rho=0.03$) between the S-value stellar activity indicator and the RV signal. $\log(\mathrm{R}'_{\mathrm{HK}}) = -5.178 \pm 0.052$, consistent with low stellar activity. We also note that a LS periodogram of the RV data suggests there is no significant periodicity remaining after subtracting a best-fit with the three known planets. We inject additional planets on circular orbits, and recover them with an LS periodogram. We find that planets with a $K$ amplitude $>$7 m s$^{-1}$ are ruled out at the 2-$\sigma$ level with periods from 55 to 1800 days. This corresponds to a 0.5 $\Mjup$ planet in a 5 year orbit, with more stringent mass constraints for shorter periods. Any proposed mechanism for the excitation of eccentricity of planet e to its observed value, must account for this constraint.

\subsection{Photometry and RV Simultaneous Fit}

We next perform a photodynamic fit including both the HIRES RV data and all quarters of \Kepler photometry data detrended as described in \S\ref{sec:methods}. We include all 18 quarters of \Kepler data including the 15 containing short cadence (58 second integration) data, which is vital for resolving transit ingress and egress and thus precise transit times. A segment of the long cadence lightcurve with transits highlighted is shown in Fig.~\ref{fig:k65lc}. Planets c and d are near a 7:5 resonance with potential TTVs of $\sim$5 minutes assuming modest eccentricities ($e\lesssim0.05$) as calculated using the formalism in \citet{2016ApJ...821...96D}. 

\begin{figure}
\centerline{
\includegraphics[scale=0.55]{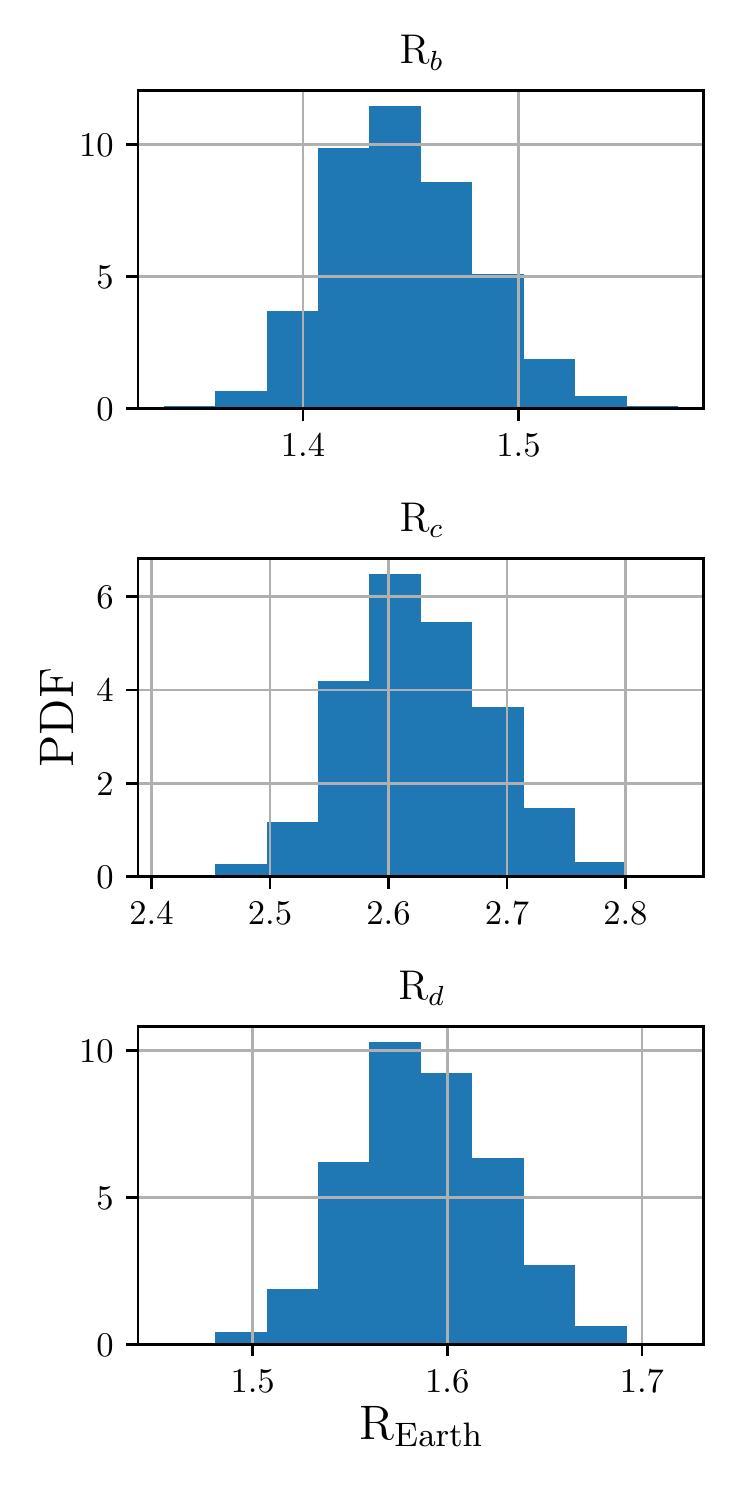}
\includegraphics[scale=0.55]{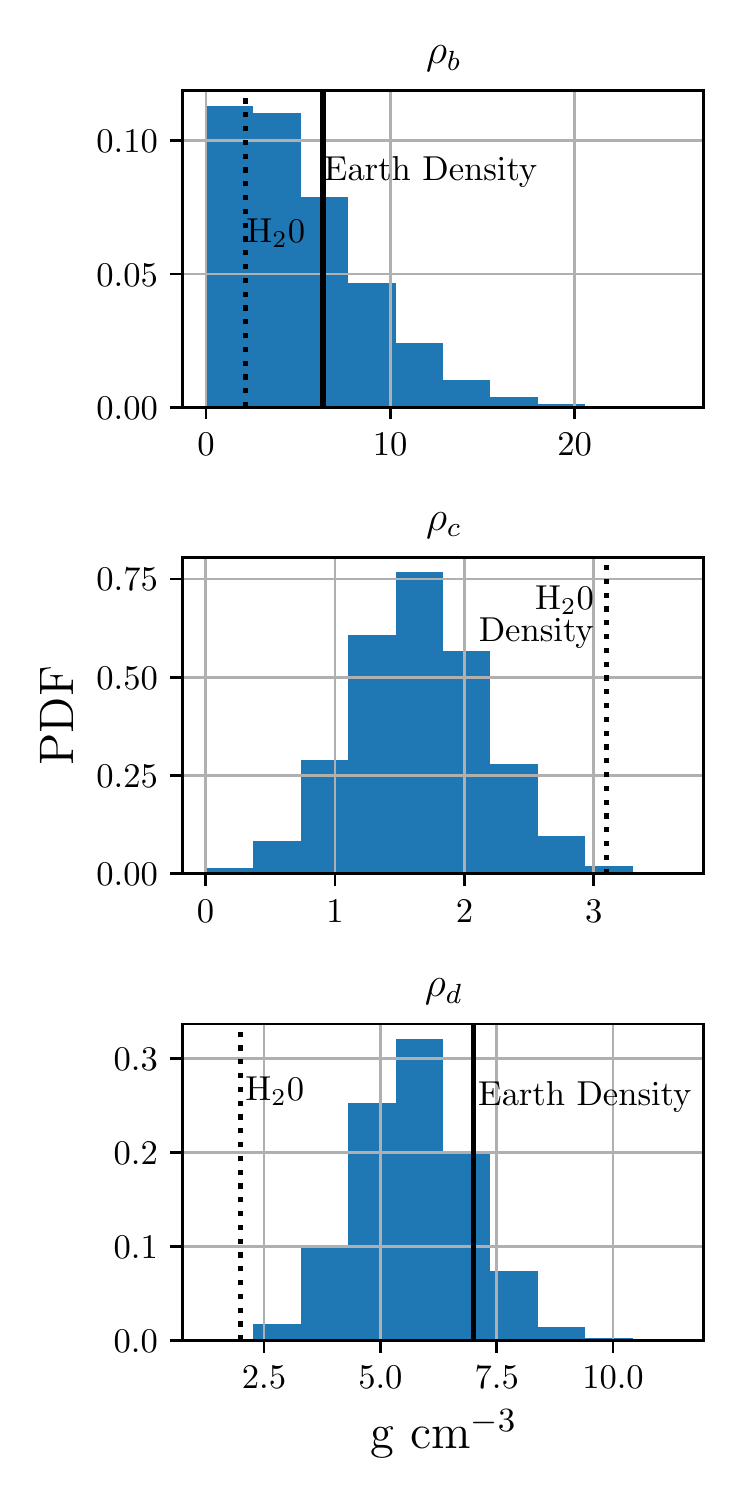}
}
\caption{ Radius and density posteriors of the transiting planets in the Kepler-65 system from the joint RV-\Kepler photometry fit. The theoretical density of an Earth-composition planet of the observed radius is shown as a solid vertical line \citep{2015ApJ...800..135D}. A dotted vertical line indicates the density of a 100\% water planet, essentially indicating lower bound on the density of a planet without an H/He atmosphere contributing significantly to the observed radius. Since planet c lies entirely below the H$_2$O density, it must have a significant H/He envelope, while planet d's allowed densities are consistent with a wider range of compositions. \label{fig:k65rrho}
}
\end{figure}

Our model follows that described in \S\ref{sec:ttvmethods}. Additionally, we fix $\Omega=0$ for the inner three planets since their mutual inclinations are not well-constrained by the data but the planets are likely nearly coplanar due the fact that all three transit. We allow $\Omega_e$ to vary to test if the data constrain the mutual inclination between the compact inner planets and the outer giant planet. This choice also allows the posteriors of all other fitted parameters to marginalize over the uncertainty in $\Omega_e$. We fix $R_e=0.01R_\star$ since it is unconstrained by the data, except that the planet does not produce significant transits. We apply a half Gaussian $e$ prior on the inner 3 planets with $\sigma_e=0.05$. This is justified since compact multiply transiting systems of small planets have low eccentricities \citep{2015ApJ...808..126V,2016PNAS..11311431X}. On the other hand, the giant planet is given a uniform $e$ prior due to the wide range of eccentricities in massive long-period planets. We apply a $\sin i$ geometric prior on the planetary inclinations. Results of a 60 chain differential evolution Markov chain Monte Carlo (DEMCMC) simulation \citep{TerBraak2005} run for 300,000 generations after burn-in are shown in Figs.~\ref{fig:k65tranfold}, \ref{fig:k65emass}, \ref{fig:k65masses}, and \ref{fig:k65rrho} and summarized in Table~\ref{table:k65}. This DEMCMC was stopped when the Gelman-Rubin statistic was $<1.2$ for all parameters and the chains remained stationary, indicating no upward or downward trends with time and no spreading (i.e., the parameter distributions for the first 150,000 generations are similar to the final 150,000 generations). Fig.~\ref{fig:65corner} illustrates the degeneracies between the derived masses and eccentricities of the planets via a corner plot of the appropriately transformed DEMCMC parameters.

\begin{figure}
\centerline{
\includegraphics[scale=0.5]{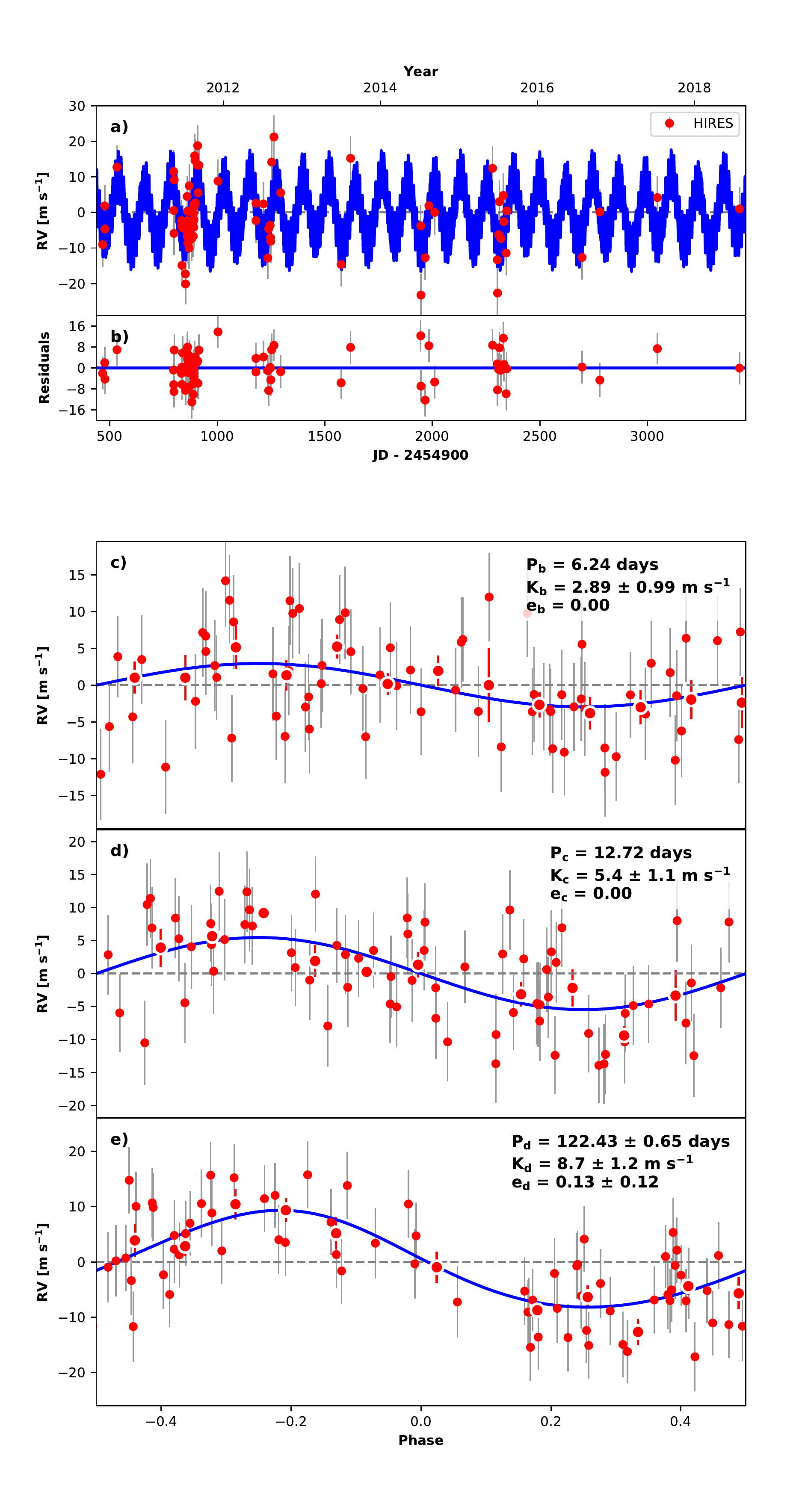}
}
\caption{\emph{Top Panel:}  Kepler-25 RV \texttt{radvel} best-fit model and residuals. \emph{Lower Panels:} RV data for each planet phase-folded at the best-fit orbital period with all other planet's signal removed.\label{fig:k25rvbest}
}
\end{figure}

\begin{deluxetable*}{llrrrrrrr}
\tablecaption{Kepler-25 Model Comparison\label{table:k25mc}}
\tablehead{\colhead{AICc Qualitative Comparison} & \colhead{Free Parameters} & \colhead{$N_{\rm free}$$^a$} & \colhead{$N_{\rm data}$} & \colhead{RMS} & \colhead{$\ln{\mathcal{L}}$} & \colhead{BIC} & \colhead{AICc} & \colhead{$\Delta$AICc}}
\startdata
  AICc Favored Model & $K_{b}$, $K_{c}$, $K_{d}$, $\sigma_\mathrm{jitter}$, $\gamma$ & 7 & 71 & 6.14 & -235.90 & 501.64 & 487.58 & 0.00 \\
  \hline 
   Strongly Disfavored & $K_{b}$, $K_{c}$, $e_{d}$, $K_{d}$, $\sigma_\mathrm{jitter}$,$\gamma$ & 9 & 71 & 6.10 & -235.51 & 509.38 & 491.97 & 4.39 \\
   & $K_{c}$, $K_{d}$, $\sigma_\mathrm{jitter}$, $\gamma$ & 6 & 71 & 6.55 & -240.51 & 506.60 & 494.34 & 6.76 \\
  \hline 
  Ruled Out    & $K_{b}$, $K_{d}$, $\sigma_\mathrm{jitter}$, $\gamma$& 6 & 71 & 7.28 & -248.42 & 522.41 & 510.14 & 22.56 \\
  & $K_{b}$, $K_{c}$, $\sigma_\mathrm{jitter}$, $\gamma$& 4 & 71 & 8.65 & -260.70 & 538.44 & 530.00 & 42.42 \\
\enddata
\tablenotetext{a}{$P_d$ and $T_{0,d}$ are allowed to vary whenever $K_d$ is a free parameter. Each $e$ also encodes two free parameters: $\sqrt{e} \cos\omega$ and $\sqrt{e} \sin \omega$.
}
\end{deluxetable*}

\subsection{Kepler-65 Discussion}
The radii of planets b, c, and d span the observed short period exoplanet radius gap, a minimum in the planetary radius distribution between that occurs at $\approx 1.8 R_\oplus$ \citep{2017AJ....154..109F}. This gap is thought to divide high-density super-Earths and low-density sub-Neptunes due to photo-evaporation \citep{2013ApJ...775..105O,2017ApJ...847...29O}. Planet c's relatively large radius ($R = 2.623^{+0.066}_{-0.056} R_\oplus$) combined with its low density ($\rho_c =1.64 ^{+0.53}_{-0.51}$ g cm$^{-3}$; see Fig.~\ref{fig:k65rrho}) suggests it may have been able to maintain a significant fraction of its primordial gas envelope. 
The density and uncertainties ($\rho_d = 5.7 \pm 1.2$ g cm$^{-3}$) of planet d are consistent with it either having a rocky composition or possessing H/He envelope despite its small radius ($R_d  = 1.587 \pm 0.04 R_\oplus$). The former is preferred by photo-evaporation models which suggest the planet should be rocky since it receives $\sim$400 times the Earth's incident flux \citep{2018MNRAS.479.5303L}. For comparison, an Earth-composition planet of planet d's radius is expected to have a density of 7 g cm$^{-3}$ \citep{2015ApJ...800..135D}. 
\citet{2013ApJ...775..105O} demonstrate that for solar-like stars, core masses of $\sim$6-8$M_\oplus$ at planet's c and d's semi-major axes (0.69 AU and 0.85 AU respectively) undergo a transition from low-density planets with envelopes to bare cores. Although planet c receives more stellar flux than planet d, it is up to 80\% more massive at the 1-$\sigma$ level. If planet c's mass is indeed on the higher end of the derived posteriors, it is thus consistent with the photo-evaporation picture due to the strong $M^{2.4}$ dependence on the necessary flux to erode an atmosphere \citep{2013ApJ...776....2L}. Planet b's density and composition remains mostly unconstrained. Parameters of the outer giant planet e are somewhat uncertain due to the RV data alone constraining its orbital parameters, however its mass is $\gtrsim0.6\Mjup$ and its eccentricity is $0.28 \pm 0.07$, higher than most planets in multi-planet systems (see Fig.~\ref{fig:k65emass}). We note that no other giant planets ($M>0.5\Mjup$) within a factor of 5 in orbital period of planet e are present in the system, determined from the lack of additional periodic RV signals. The mutual inclination between the outer giant planet e and the inner 3 planets is is not meaningfully constrained by the data.

\begin{deluxetable}{lrr}
\tablecaption{Kepler-25 MCMC Posteriors \label{table:k25}}
\tablehead{\colhead{Parameter$^a$} & \colhead{Median$^{\mathrm{84}^{th}\mathrm{ Percentile}}_{\mathrm{16}^{th}\mathrm{ Percentile}}$} & \colhead{Unit}}
{\def\baselinestretch{0.95}
\startdata
\hline
\multicolumn{3}{c}{\bf Fitted Parameters}\\
\hline
P$_b$      	& 6.238297$^{+1.7e\mymathhyphen05}_{-1.7e\mymathhyphen05}$  &   days    \\
T$_0,$$_b$      	& 803.42004$^{+0.00012}_{-0.00011}$  &   BJD-2454900    \\
$\sqrt{e}\cos \omega$$_b$	& 0.042$^{+0.017}_{-0.036}$  &       \\
$\sqrt{e}\sin \omega$$_b$	& 0.007$^{+0.038}_{-0.035}$  &       \\
i$_b$           	& 92.827$^{+0.084}_{-0.083}$  &    degrees   \\
M$_b$  	& 0.0275$^{+0.0079}_{-0.0073}$  &    $\Mjup$   \\
R$_p$/R$_\star$$_b$ 	& 0.01916$^{+5.1e\mymathhyphen05}_{-4.8e\mymathhyphen05}$  &       \\
P$_c$      	& 12.7207$^{+0.00011}_{-0.0001}$  &    days      \\
T$_0,$$_c$      	& 811.15013$^{+0.00014}_{-0.00014}$  &     BJD-2454900   \\
$\sqrt{e}\cos \omega$$_c$	& -0.024$^{+0.067}_{-0.053}$  &       \\
$\sqrt{e}\sin \omega$$_c$	& 0.004$^{+0.065}_{-0.062}$  &       \\
i$_c$           	& 92.764$^{+0.042}_{-0.039}$  &   degrees    \\
$\Omega$$_c$    	& -0.45$^{+0.19}_{-0.25}$  &   degrees    \\
M$_c$  	& 0.0479$^{+0.0041}_{-0.0051}$  &    $\Mjup$    \\
R$_p$/R$_\star$$_c$ 	& 0.03637$^{+0.00012}_{-0.00012}$  &       \\
P$_d$      	& 122.4$^{+0.8}_{-0.71}$  &     days     \\
T$_0,$$_d$      	& 815.0$^{+6.8}_{-7.2}$  &     BJD-2454900   \\
$\sqrt{e}\cos \omega$$_d$	& 0.07$^{+0.27}_{-0.29}$  &       \\
$\sqrt{e}\sin \omega$$_d$	& 0.16$^{+0.23}_{-0.28}$  &       \\
M$_d$  	& 0.226$^{+0.031}_{-0.031}$  &   $\Mjup$     \\
M$_\star$           	& 1.165$^{+0.027}_{-0.029}$  &   $M_\odot$    \\
R$_\star$           	& 1.316$^{+0.016}_{-0.015}$  &    $R_\odot$    \\
c$_1$           	& 0.351$^{+0.049}_{-0.051}$  &       \\
c$_2$           	& 0.198$^{+0.056}_{-0.055}$  &       \\
$\sigma_\mathrm{jitter}$    	& 5.44$^{+0.74}_{-0.65}$  &  m s$^-1$     \\
$\gamma$  & $1.67^{+0.85}_{-0.88}$  &  m s$^-1$     \\
\hline
\multicolumn{3}{c}{\bf Derived Parameters}       \\
\hline
i$_c$/i$_b$     	& 0.99932$^{+0.0005}_{-0.00048}$  &       \\
M$_b$           	& 8.7$^{+2.5}_{-2.3}$  &     $M_\oplus$  \\
M$_c$           	& 15.2$^{+1.3}_{-1.6}$  &    $M_\oplus$   \\
M$_d$           	& 71.9$^{+9.8}_{-9.8}$  &    $M_\oplus$   \\
$M_d\, \sin i_d$	& 0.226$^{+0.031}_{-0.031}$  &  $\Mjup$     \\
R$_b$           	& 2.748$^{+0.038}_{-0.035}$  &    $R_\oplus$   \\
R$_c$           	& 5.217$^{+0.07}_{-0.065}$  &    $R_\oplus$   \\
$\rho$$_b$      	& 2.32$^{+0.67}_{-0.61}$  &    g cm$^{-3}$   \\
$\rho$$_c$      	& 0.588$^{+0.053}_{-0.061}$  &     g cm$^{-3}$    \\
$e_b$           	& 0.0029$^{+0.0023}_{-0.0017}$  &       \\
$e_c$           	& 0.0061$^{+0.0049}_{-0.0041}$  &       \\
$e_d$           	& 0.13$^{+0.13}_{-0.09}$  &       \\
\enddata
}
\tablenotetext{a}{Osculating orbital elements are valid at $T_\mathrm{epoch} = 2,455,700$ BJD.
}
\end{deluxetable}

\section{Kepler-25}

Kepler-25 (KOI-244, KIC 4349452) is a system of two transiting planets between 2 and 5 $R_\oplus$ and with orbital periods near a 2:1 ratio at 6.2 and 12.7 days \citep{2012MNRAS.421.2342S}. \citet{2014ApJS..210...20M} use RVs to identify an additional non-transiting $\sim90M_\oplus$ companion on a 123 day orbit. We use additional HIRES RV observations and an extended time baseline to more precisely characterize both the inner transiting planets and the outer non-transiting planet. The inner planets both exhibit TTVs \citep{2012MNRAS.421.2342S,2016ApJS..225....9H}. The TTVs were used to estimate the planetary masses by \citet{2017AJ....154....5H} who found $M_b = 4^{+4}_{-2} M_\oplus$ and $M_c = 10^{+3.5}_{-2.5} M_\oplus$ subject to a high mass prior and $M_b = 0.4^{+1.5}_{-0.2} M_\oplus$ and $M_c =1.4^{+4}_{-0.6} M_\oplus$ with a less informative prior. By combining the RV data and the photometric data simultaneously, we put more precise mass and density constraints on the inner planets. 

\citet{2013ApJ...771...11A} demonstrate that planet c's orbit is nearly aligned with the host star rotation (i.e., it has obliquity $\lesssim 22^\circ$) via a 2-night Rossiter-McLaughlin (R-M) measurement. For our RV fit, we include HIRES RVs made as part of the CPS survey \citep{2010ApJ...721.1467H} and choose to include only two RV measurements of the system on the nights of the R-M observations (before and after each transit). Hence the R-M effect does not bias our fit, and any correlated noise on the $\sim12$ hour timescale of the observations does not get unfairly favored by the density of observations on these nights. The complete list of the RV data used for this fit is given in Table~\ref{table:k25rvs}.  

The dynamics of the inner transiting planets of Kepler-25 were also studied in detail by \citet{2018MNRAS.480.1767M}. They stated that the system is consistent with either being in a periodic (resonant) configuration or not within the uncertainties in the planets' eccentricities. Although somewhat far from resonance with a period ratio of 2.039, the divergence of the resonance width at low eccentricity permits a periodic solution for $e \lesssim 0.002$ and relatively high planetary mass to reproduce the observed TTV amplitudes. The addition of our RV data set reduces the mass-eccentricity degeneracy found in the TTVs alone.

\subsection{Kepler-25 RV Only Fit}

We first perform a Keplerian fit with \texttt{radvel} on the HIRES RV data set alone (Table~\ref{table:k25rvs}). The two inner planets have well determined orbital periods and phases from \emph{Kepler}, which we fix at the observed values. As with Kepler-65, we set $e=0$ and allow all parameters of the newly discovered outer planet to vary. Our best fit model is shown in Fig.~\ref{fig:k25rvbest}. All three planets are detected at high significance (Table~\ref{table:k25mc}). We also note that eccentricity is not confidently detected in planet d since including $e_d$ is disfavored by the AICc \citep[small-sample corrected Akaike Information Criteria;][]{burnham2003model}, and is constrained to $e_d<0.28$ at the 1-$\sigma$ level if included\footnote{This model comparison feature is freely available as part of the \texttt{radvel} package (https://github.com/California-Planet-Search/radvel).}.  

\begin{figure}
\centerline{
\includegraphics[scale=0.45]{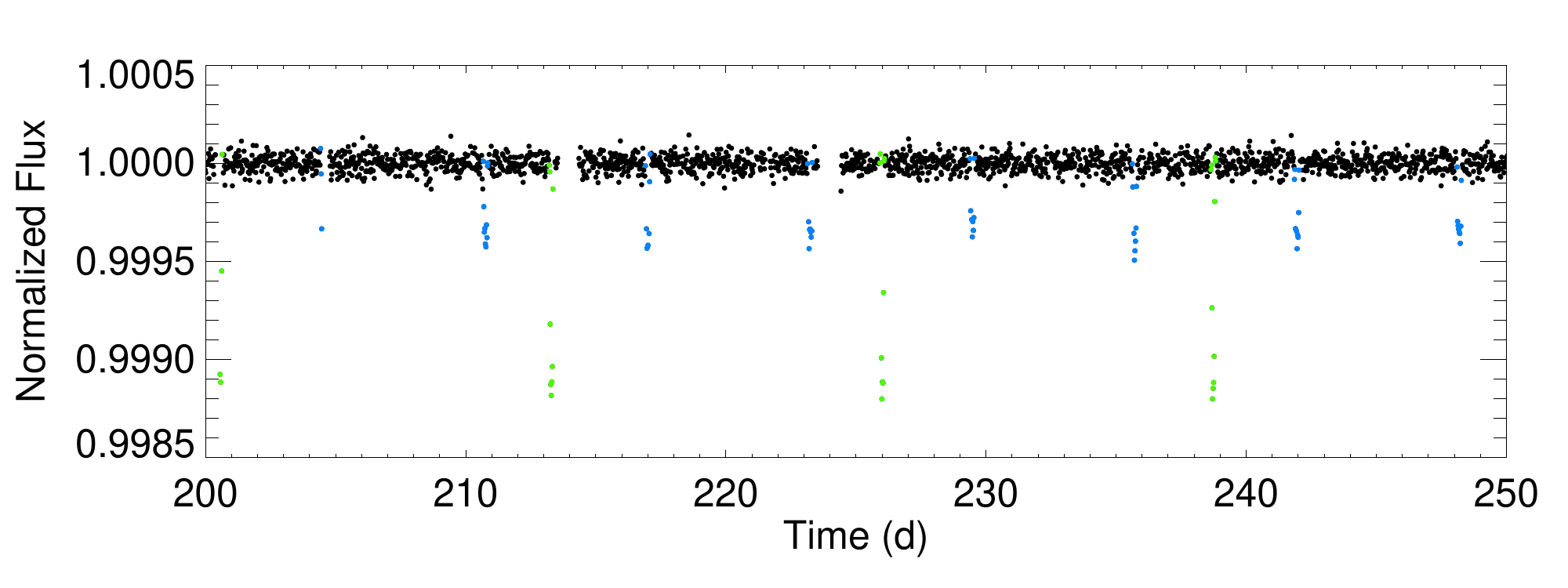}
}
\caption{A segment of the long-cadence portion of Kepler-25's detrended lightcurve (BJD-2454900) with transit events highlighted in blue (planet b) and green (planet c). \label{fig:k25lc}
}
\end{figure}

\begin{figure}
\centerline{
\includegraphics[scale=0.6]{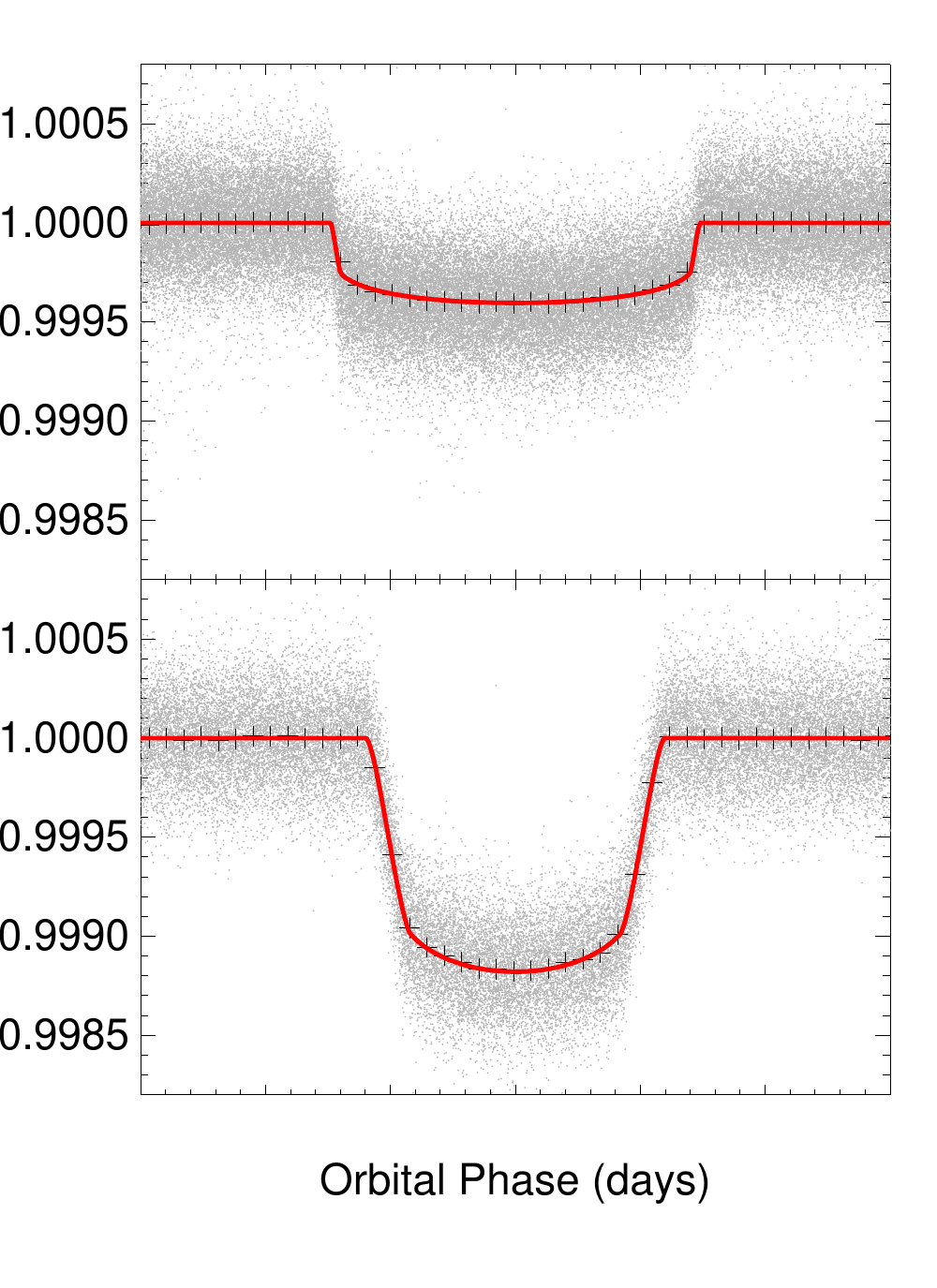}
}
\caption{ Kepler-25's planets b (top) and c (bottom) phase-folded with TTVs removed. Gray points are individual \Kepler data, black crosses are data binned in 10 minute intervals, and the red line is a best fit transit model. \label{fig:k25tranfold}
}
\end{figure}

\begin{figure}
\centerline{
\includegraphics[scale=0.6]{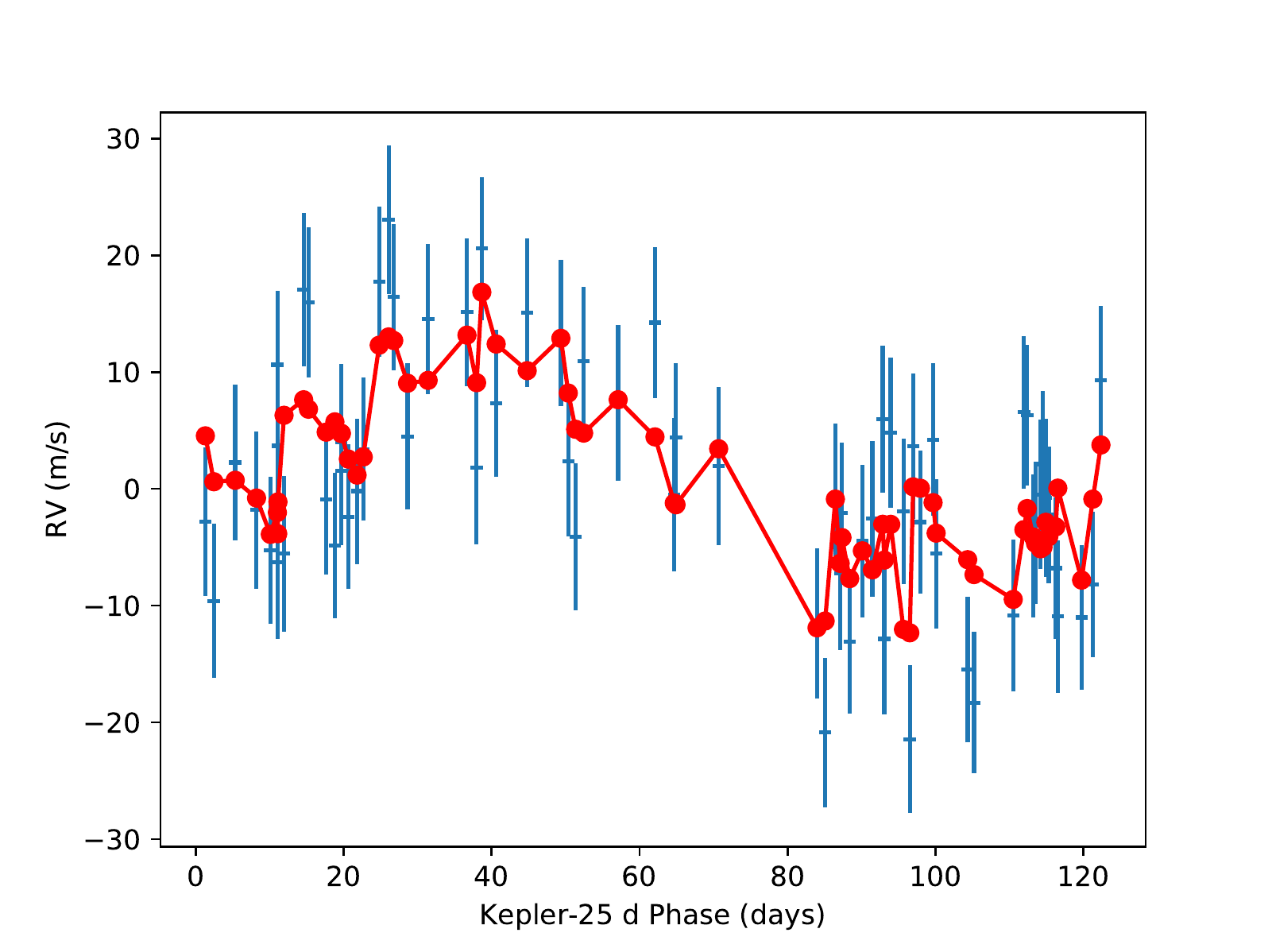}
}
\caption{ The best-fit solution's theoretical RVs (red) and the HIRES RV data (crosses with uncertainties including the best-fit RV jitter of 5.3 m s$^{-1}$) phased at the best-fit orbital period of the long-period giant planet (planet d) in Kepler-25. The small amplitude variation in theoretical points is due to the RV contribution of the two inner planets. Their low amplitudes and correlation with variations in the data agree with the fit using only RVs. 
\label{fig:k25xmass}
}
\end{figure}

\begin{figure}
\centerline{
\includegraphics[scale=0.6]{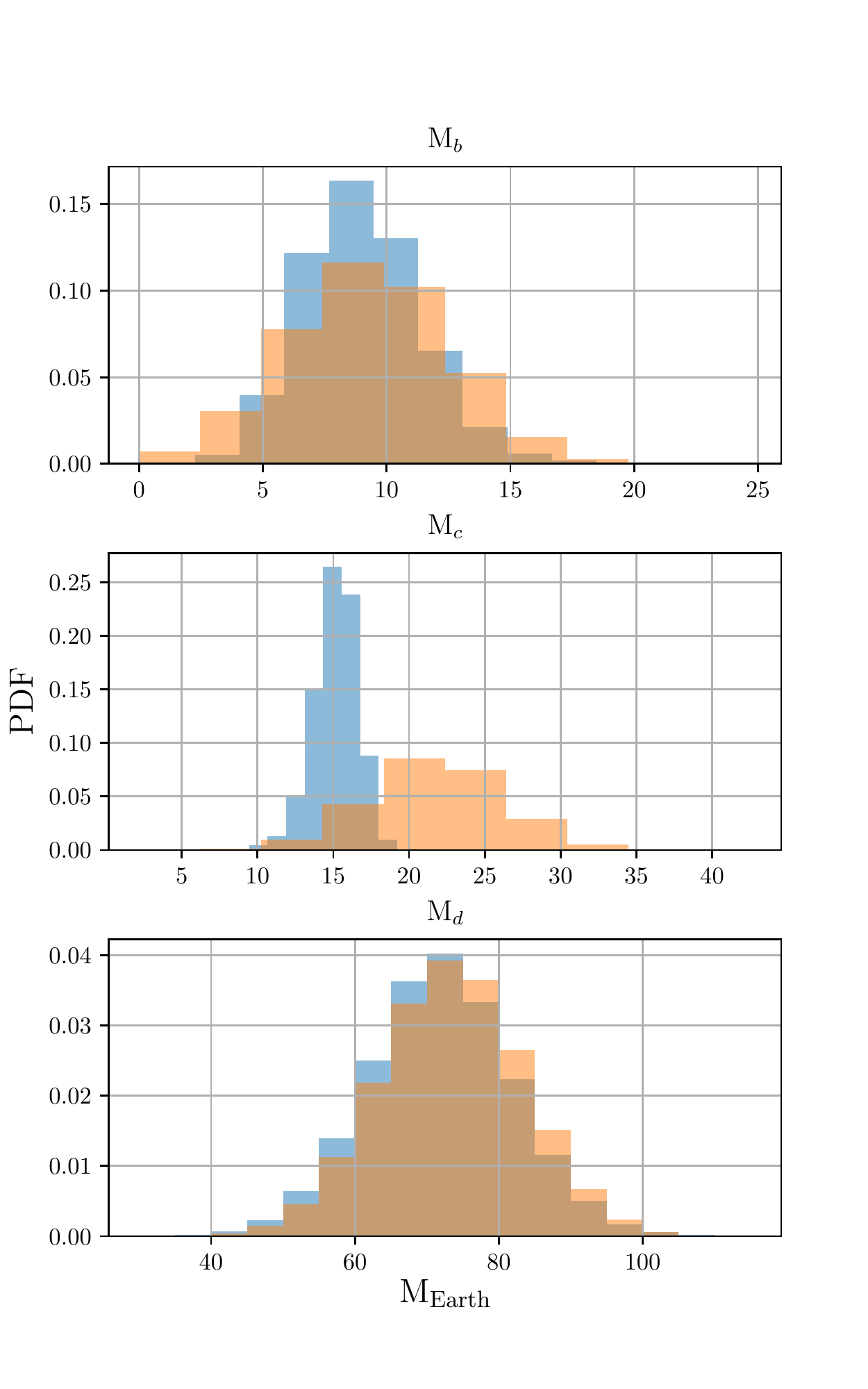}
}
\caption{Mass posteriors for all planets in Kepler-25 from the joint RV-\Kepler photometry fits (blue) and the RVs alone (orange). The precision of planet c's mass increases dramatically with the addition of the TTV information, and the RVs indicate that the mass is at the higher end of the distribution allowed by the TTVs alone \citep{2018MNRAS.480.1767M}. \label{fig:k25masses}
}
\end{figure}

In addition to the most significant peak at 123 days in an LS Periodogram of the RVs, we note a secondary peak at 91 days with less power. A model comparison of a Keplerian RV fit with the outer planet at a 90 day orbital period compared to a 123 day period fit has a $\Delta$ Bayesian Information Criterion (BIC) $=6$ (that is, a log likelihood difference of 4 for 7 free parameters), providing substantial evidence against the 91 day period, but not conclusively ruling it out. We emphasize that the RV posteriors for planets b and c are statistically similar between fits with the outer giant at a 90 days and a 123 days---the mass posteriors for both b and c are consistent to $<0.5 \sigma$ between the fits. Therefore the interpretation of the inner planets properties would not qualitatively change, even if the outer giant planet is indeed at the disfavored 91 day orbital period. 

We also consider stellar activity as a potential false positive for an apparent planetary signal. A periodogram of the S-values of the HIRES RV data does not show a peak at the period of the putative outer planet. $\log(\mathrm{R}'_{\mathrm{HK}}) = -5.21 \pm 0.15$, consistent with low stellar activity.  We also note that a LS periodogram suggests there is no significant periodicity remaining after subtracting a best-fit with the three known planets. We attempt to inject an additional planet on a circular orbits, and recover it with an LS periodogram. We find that planets with a $K$ amplitude $>$7 m s$^{-1}$ are ruled out at the 2-$\sigma$ level with periods from approximately 60 to 3000 days.

\subsection{Photometry and RV Simultaneous Fit}

The TTVs detected between the two transiting planets \citep{2017AJ....154....5H,2016ApJS..225....9H} present a degeneracy between their masses and eccentricities \citep{2012ApJ...761..122L,2018MNRAS.480.1767M}. Including the RV data may help break this degeneracy as the RVs can put bounds on the allowed masses. Therefore we perform a photodynamic fit including both the HIRES RV data simultaneously and all quarters of \Kepler photometry data (a segment of which is shown in Fig.~\ref{fig:k25lc}).

\begin{figure}
\centerline{
\includegraphics[scale=0.45]{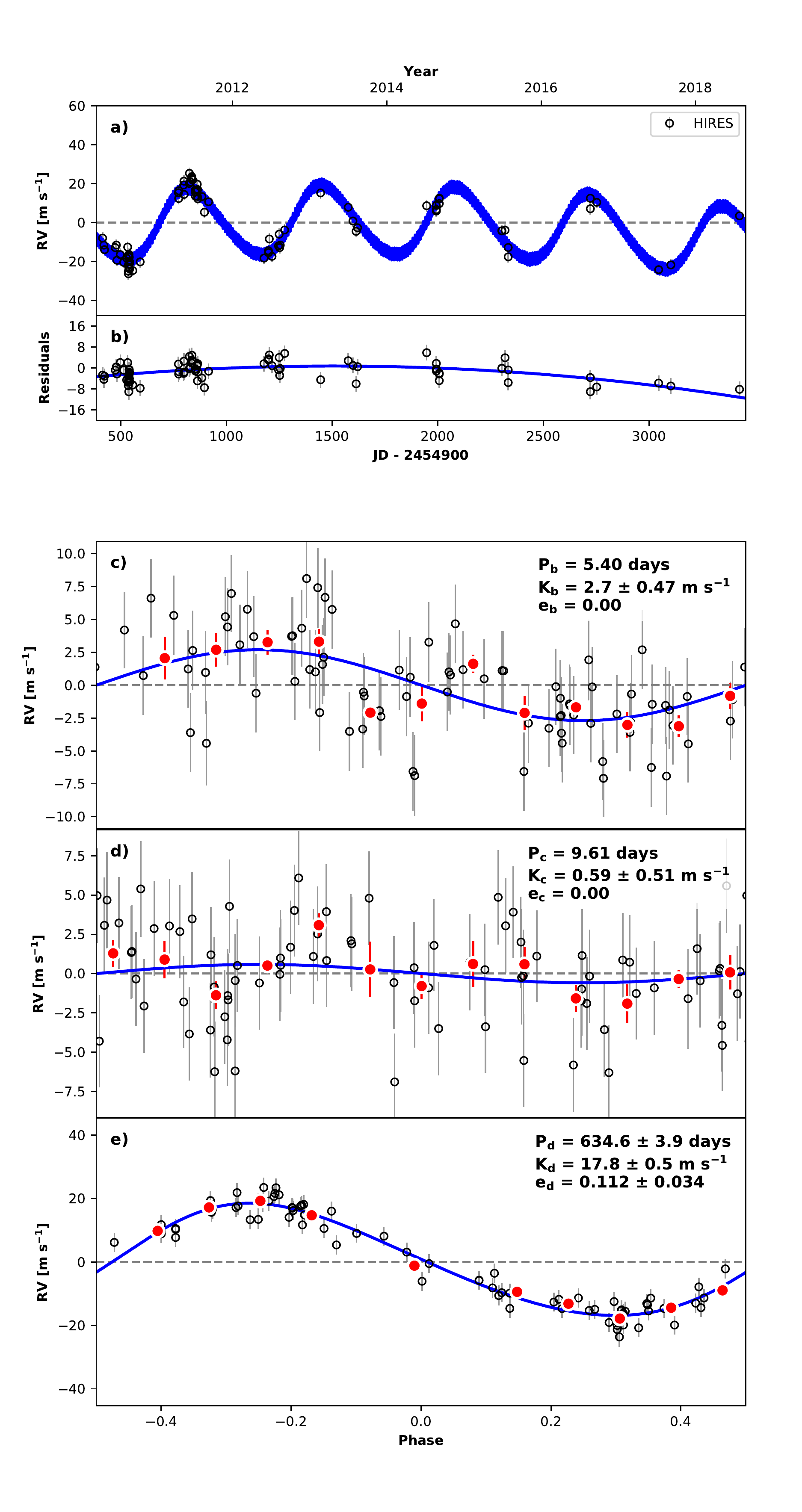}
}
\caption{\emph{Top Panel:} Kepler-68 RV \texttt{radvel} best-fit model and residuals. \emph{Lower Panels:} RV data for each planet phase-folded at the best-fit orbital period with all other planet's signal removed.\label{fig:k68rvbest}
}
\end{figure}

\begin{deluxetable}{lrr}
\tablecaption{Kepler-68 \texttt{radvel} MCMC Posteriors\label{table:k68params}}
\tablehead{
  \colhead{Parameter} & 
  \colhead{Credible Interval} & 
  \colhead{Units}
}
\startdata
\hline
\sidehead{\bf{Orbital Parameters}}
 $P_{b}$ & $\equiv5.3988$  & days \\
  $T_{0,b}$ & $\equiv803.2978$ & BJD-2454900 \\
  $e_{b}$ & $\equiv0.0$  &  \\
  $\omega_{b}$ & $\equiv0.0$ & radians \\
  $K_{b}$ & $2.7^{+0.48}_{-0.46}$ &  m s$^{-1}$ \\
  $P_{c}$ & $\equiv9.6051$ & days \\
  $T_{0,c}$ & $\equiv808.9682$ & BJD-2454900 \\
  $e_{c}$ & $\equiv0.0$ &   \\
  $\omega_{c}$ & $\equiv0.0$ & radians \\
  $K_{c}$ & $0.59^{+0.50}_{-0.52}$  & m s$^{-1}$ \\
  $P_{d}$ & $634.6^{+4.1}_{-3.7}$ & days \\
  $T_{0,d}$ & $978\pm 11$  & BJD-2454900 \\
  $e_{d}$ & $0.112^{+0.035}_{-0.034}$  &  \\
  $\omega_{d}$ & $-1.13^{+0.36}_{-0.45}$ & radians \\
  $K_{d}$ & $17.75^{+0.50}_{-0.49}$ & m s$^{-1}$ \\
\hline
\sidehead{\bf{Other Parameters$^a$}}
  $\gamma_{\rm HIRES}$ & $1.85^{+0.74}_{-0.72}$  &  \\
  $\dot{\gamma}$ & $-0.00319^{+0.00075}_{-0.00078}$  & m s$^{-1}$ d$^{-1}$ \\
  $\ddot{\gamma}$ & $-3.23e\mymathhyphen06^{+8.6e\mymathhyphen07}_{-8.3e\mymathhyphen07}$  & m s$^{-1}$ d$^{-2}$ \\
  $\sigma_\mathrm{jitter}$ & $2.93^{+0.32}_{-0.29}$ & m s$^-1$  \\
  \hline
\sidehead{\bf{Derived Parameters$^b$}}
  $M_b \sin i$ & $7.65^{+1.37}_{-1.32}$ & $M_\oplus$  \\
  $M_c \sin i $ & $2.04^{+1.72}_{-1.78}$ & $M_\oplus$ \\
  $M_d \sin i $ & $0.77^{+0.03}_{-0.03}$ & $\Mjup$ \\
\enddata
\tablenotetext{a}{
  Reference epoch for $\gamma$,$\dot{\gamma}$,$\ddot{\gamma}$: 2000 
}
\tablenotetext{b}{
  With system scale set using stellar data from \citep{2018AJ....156..264F}.
}

\end{deluxetable}

\begin{deluxetable*}{llrrrrrrr}
\tablecaption{Kepler-68 Model Comparisons\label{table:k68mc}}
\tablehead{\colhead{AICc Qualitative Comparison} & \colhead{Free Parameters$^a$} & \colhead{$N_{\rm free}$$^b$} & \colhead{$N_{\rm data}$} & \colhead{RMS} & \colhead{$\ln{\mathcal{L}}$} & \colhead{BIC} & \colhead{AICc} & \colhead{$\Delta$AICc}}
\startdata
  AICc Favored Model & $K_{b}$, $e_{d}$, $K_{d}$, $\dot{\gamma}$, $\ddot{\gamma}$, $\sigma_\mathrm{jitter}$, $\gamma$ & 10 & 82 & 3.05 & -224.92 & 493.91 & 472.94 & 0.00 \\
  \hline 
  Nearly Indistinguishable & $K_{b}$, $K_{c}$, $e_{d}$, $K_{d}$, $\dot{\gamma}$, $\ddot{\gamma}$, $\sigma_\mathrm{jitter}$, $\gamma$ & 11 & 82 & 3.03 & -224.19 & 496.86 & 474.16 & 1.22 \\
   \hline
  Strongly Disfavored & $K_{b}$, $K_{d}$, $\dot{\gamma}$, $\ddot{\gamma}$, $\sigma_\mathrm{jitter}$, $\gamma$ & 8 & 82 & 3.29 & -231.15 & 497.55 & 480.27 & 7.33 \\
     & $K_{b}$, $K_{c}$, $K_{d}$, $\dot{\gamma}$, $\ddot{\gamma}$, $\sigma_\mathrm{jitter}$, $\gamma$ & 9 & 82 & 3.27 & -230.60 & 500.86 & 481.70 & 8.76 \\
  \hline 
  Ruled Out & $K_{b}$, $e_{d}$, $K_{d}$, $\dot{\gamma}$, $\sigma_\mathrm{jitter}$, $\gamma$ & 9 & 82 & 3.33 & -231.99 & 503.64 & 484.48 & 11.54 \\
     & $K_{b}$, $K_{c}$, $e_{d}$, $K_{d}$, $\dot{\gamma}$, $\sigma_\mathrm{jitter}$, $\gamma$ & 10 & 82 & 3.32 & -231.68 & 507.43 & 486.46 & 13.52 \\
     & $K_{b}$, $e_{d}$, $K_{d}$, $\sigma_\mathrm{jitter}$, $\gamma$ & 8 & 82 & 3.44 & -234.90 & 505.06 & 487.78 & 14.84 \\
   & $K_{b}$, $K_{c}$, $e_{d}$, $K_{d}$, $\sigma_\mathrm{jitter}$, $\gamma$ & 9 & 82 & 3.44 & -234.73 & 509.13 & 489.97 & 17.03 \\
\enddata
\tablenotetext{a}{$P_d$ and $T_{0,d}$ are allowed to vary whenever $K_d$ is a free parameter. Each $e$ also encodes two free parameters: $\sqrt{e} \cos\omega$ and $\sqrt{e} \sin \omega$.
}
\end{deluxetable*}

We fix $\Omega_b=0$, but allow $\Omega_c$ to vary since we expect the inner two planets to be tightly coupled and therefore have mutual inclination well-constrained by the TTVs and lack of large duration variations (TDVs). We also fix $R_d=0.01R_\star$ since it is completely unconstrained by the data and does not affect our fits except that the planet should not produce significant transits since none are observed. 

Similar to the Kepler-65 fit, we apply a Gaussian $e$ prior on the inner two planets with $\sigma_e=0.05$, and also note that Kepler-25's planets must have eccentricities well within that range for any physical masses \citep{2018MNRAS.480.1767M}. On the other hand, the giant planet has a uniform $e$ prior due to the wide range of eccentricities in massive long-period planets. We again apply stellar mass and radius data from \citet{2018AJ....156..264F} as priors to get absolute radius and density information for the system. The planetary masses and other parameters are all given a uniform prior. 

We consider allowing $\Omega_d$ and $i_d$ to vary, but find that our fits slightly favor a highly mutually inclined configuration ($I \gtrsim 30^\circ$) under a uniform $\Omega$ prior and a geometric $\sin i$ prior. However the improvement in the likelihood is small, compared to the introduction of free parameters. We compare the models using the BIC (Bayesian Information Criteria) and AICc. Specifically, the $\Delta$BIC $=$ -13 favors the extra free parameters while the $\Delta$AIC $=$ 12 disfavors them, suggesting that it is not clear if the addition of the inclination parameters is well-founded in our model given the data. Therefore in our final fit reported here, we fix $\Omega_d = \Omega_b = 0$ and $i_d = 92^\circ \approx i_b \approx i_c$, since the data may be insufficient to determine mutual inclination and most known multiplanet systems are nearly coplanar \citep{2014ApJ...790..146F,2018ApJ...860..101Z}. We expect this choice to have little effect on the posteriors of other parameters in our model and will prevent misleading high mutual inclination fits from dominating the reported posteriors.

Results of a 40 chain DEMCMC \citep{TerBraak2005} run for 220,000 generations after burn-in are shown in Figs.~\ref{fig:k25tranfold}, \ref{fig:k25xmass}, and \ref{fig:k25masses} and summarized in Table~\ref{table:k25}. This MCMC was stopped when the Gelman-Rubin statistic was $<1.2$ for all parameters and the chains remained stationary, indicating no upward or downward trends with time and no spreading (i.e., the parameter distributions for the first 110,000 generations are similar to the final 110,000 generations). Fig.~\ref{fig:25corner} illustrates the degeneracies between the derived masses and eccentricities of the planets via a corner plot of the appropriately transformed DEMCMC parameters.

\subsection{Kepler-25 Discussion}
Our RV-TTV combined results favor high masses and low eccentricities (see Fig.~\ref{fig:k25masses}) compared to the range of possible values as explored in \citet{2018MNRAS.480.1767M}. This is primarily due to the RV signal, whose relatively large amplitudes partially break the TTV mass-eccentricity degeneracy. Our results are within 1-2$\sigma$ of the \citet{2017AJ....154....5H} result when a high mass prior was used. The low eccentricities found are consistent with the low-eccentricity resonant state that \citet{2018MNRAS.480.1767M} speculate Kepler-25 occupies; however, our results do not strictly demand resonance.

\section{Kepler-68}

The two transiting planets detected by \Kepler and one non-transiting planet detected via RVs in Kepler-68 (KOI-246, KIC 11295426) were validated by \cite{2013ApJ...766...40G}. The non-transiting outer planet was found by \citet{2014ApJS..210...20M}, who report $P_d=625\pm16$ days and $M_d \sin i = 0.84\pm0.05\Mjup$. Here we provide improved constraints from additional RV data and a longer time baseline. Since the transiting planets exhibit no TTVs, we only fit the RV data (shown in full in Table~\ref{table:k68rvs}). We again run a \texttt{radvel} fit with the inner planets' periods and phases fixed based on the \Kepler data. The results of an MCMC are summarized in Table~\ref{table:k68params}, with the best fit shown in Fig.~\ref{fig:k68rvbest}. We note that the mass posteriors are symmetric, Gaussian, and have no apparent correlation with any other parameters.

We consider several different models of the RV data with and without terms for a linear trend, quadratic curvature, and eccentricity for the outer planet and compare their relative likelihoods via BIC and AICc. Our results in Table~\ref{table:k68mc} show that the inclusion of both the long-timescale curvature terms (linear and quadratic) and planet d's eccentricity are both strongly preferred. The presence of curvature in the RVs indicates the presence of another body in the system at a period $\gtrsim 10$ years and uncertain mass. This could either be a planetary or stellar companion. At the $\sim$10 year period lower limit, the signal would correspond to a planetary $M \sin i$ of $\sim$0.6$\Mjup$, however the period and mass of this body are not bounded by the measurements and could be much greater. \citet{2016MNRAS.457.2173G} report a nearby star in a lucky imaging survey for companions at a distance of 11". They conclude the companion is likely bound to Kepler-68; however, this star is at such a great distance from Kepler-68 \citep[$\approx1600$ AU sky-projected;][]{2018A&A...616A...1G} that it is not likely the cause of the observed curvature in the RVs. A circular orbit with the sky-projected distance as the semi-major axis would have $P\sim50,000$ years, which would not be detectable over our $\sim$7-year baseline.

\begin{deluxetable}{lrr}
\tablecaption{System Summaries\label{table:summary}}
\tablehead{
  \colhead{Parameter} & 
  \colhead{Credible Interval} & 
  \colhead{Units}
}
\startdata
\hline
\sidehead{\bf{Kepler-25}}
\hline
P$_b$      	&$ 6.238297 ^{ +1.7e\mymathhyphen05}_{-1.7e\mymathhyphen05}$  &  days  \\
P$_c$      	&$ 12.7207 ^{ +0.00011}_{-0.0001}$  &  days  \\
P$_d$      	&$ 122.4 ^{ +0.80}_{-0.71}$  &   days \\
M$_b$           	&$ 8.7 ^{ +2.5}_{-2.3}$  &   $M_\oplus$ \\
M$_c$           	&$ 15.2 ^{ +1.3}_{-1.6}$  &  $M_\oplus$  \\
$M_{d} \sin i_d$	&$ 0.226 ^{ +0.031}_{-0.031}$  &  $\Mjup$  \\
R$_b$           	&$ 2.748 ^{ +0.038}_{-0.035}$  &  $R_\oplus$  \\
R$_c$           	&$ 5.217 ^{ +0.07}_{-0.065}$  &   $R_\oplus$ \\
$\rho$$_b$      	&$ 2.32 ^{ +0.67}_{-0.61}$  &  g cm$^{-3}$  \\
$\rho$$_c$      	&$ 0.588 ^{ +0.053}_{-0.061}$  & g cm$^{-3}$   \\
$e_b$           	&$ 0.0029 ^{ +0.0023}_{-0.0017}$  &    \\
$e_c$           	&$ 0.0061 ^{ +0.0049}_{-0.0041}$  &    \\
$e_d$           	&$ 0.13 ^{ +0.13}_{-0.09}$  &    \\
\hline
\hline
\sidehead{\bf{Kepler-65}}
\hline
P$_b$              &$ 2.1549209^{+8.6e\mymathhyphen06}_{-7.4e\mymathhyphen06}$  &  days  \\
P$_c$              &$ 5.859697^{+9.3e\mymathhyphen05}_{-9.9e\mymathhyphen05}$  &   days \\
P$_d$              &$ 8.13167^{+0.00024}_{-0.00021}$  &  days  \\
P$_e$              &$ 258.8^{+1.5}_{-1.3}$  &  days  \\
M$_b$                   &$ 2.4^{+2.4}_{-1.6}$  &  $M_\oplus$  \\
M$_c$                   &$ 5.4^{+1.7}_{-1.7}$  &  $M_\oplus$  \\
M$_d$                   &$ 4.14^{+0.79}_{-0.8}$  &   $M_\oplus$ \\
$M_{e} \sin i_e$    &$ 0.653^{+0.056}_{-0.055}$  &  $\Mjup$  \\
R$_b$                   &$ 1.444^{+0.037}_{-0.031}$  &  $R_\oplus$  \\
R$_c$                   &$ 2.623^{+0.066}_{-0.056}$  & $R_\oplus$   \\
R$_d$                   &$ 1.587^{+0.04}_{-0.035}$  &  $R_\oplus$  \\
$\rho$$_b$              &$ 4.4^{+4.5}_{-3.0}$  & g cm$^{-3}$   \\
$\rho$$_c$              &$ 1.64^{+0.53}_{-0.51}$  &  g cm$^{-3}$  \\
$\rho$$_d$              &$ 5.7^{+1.2}_{-1.2}$  & g cm$^{-3}$   \\
$e_b$                   &$ 0.028^{+0.031}_{-0.02}$  &    \\
$e_c$                   &$ 0.02^{+0.022}_{-0.013}$  &    \\
$e_d$                   &$ 0.014^{+0.016}_{-0.01}$  &    \\
$e_e$                   &$ 0.283^{+0.064}_{-0.071}$  &    \\
\hline
\hline
\sidehead{\bf{Kepler-68}}
\hline
  $P_{b}$ & $\equiv5.3988$ & days \\
  $P_{c}$ & $\equiv9.6051$ & days \\
  $P_{d}$ & $634.6^{+4.1}_{-3.7}$  & days \\
  $M_b$ & $7.65^{+1.37}_{-1.32}$ & $M_\oplus$  \\
  $M_c $ & $2.04^{+1.72}_{-1.78}$ & $M_\oplus$ \\
  $M_d \sin i $ & $0.77^{+0.03}_{-0.03}$ & $\Mjup$ \\
  $e_{d}$ & $0.112^{+0.035}_{-0.034}$  &  \\
\enddata
\tablenotetext{}{
}
\end{deluxetable}

A periodogram of the S-values of the HIRES RV data does not show a peak at the period of the putative planet or its harmonics. The presence of some additional weak peaks in the S-value periodogram are taken into account by the model's stellar jitter term which appropriately broadens the posteriors of the derived planetary properties. There is only weak correlation between the stellar activity indicator and the RV signal ($\rho = -0.2$). This correlation is driven in part by a single point with anomalously low S-value which provides a large lever when fitting a linear correlation which we suspect is due to sky background contamination, so we do not expect our results to be significantly biased by stellar activity. Additionally, $\log(\mathrm{R}'_{\mathrm{HK}}) = -5.153 \pm 0.037$, consistent with low stellar activity.

\section{Summary and Discussion}

In this paper we expand the sample of known long-period giant planet companions to compact, coplanar, multiply-transiting systems with the introduction of the $M\sin i = 0.65\pm0.06 \Mjup$ Kepler-65 e. Additionally, we provide improved mass and orbital element constraints on the long-period Jupiters in Kepler-25 and Kepler-68, while also tentatively suggesting the presence of a second long-period companion in Kepler-68 with $P\gtrsim10$ years. Important system parameters are summarized in Table~\ref{table:summary}. The non-transiting giant planets in all 3 systems are dynamically separated from the inner SEASNs due to their large orbital period ratios (approximately 9.6, 32, and 66 in Kepler-25, Kepler-65, and Kepler-68 respectively). Our joint RV-TTV analysis also points to high masses and density for the transiting planets of Kepler-25, consistent with previous work suggesting they may be in a low-eccentricity, periodic configuration \citep{2018MNRAS.480.1767M}. 

We also consider the possibility that the observed long-period, eccentric giant planets are actually a pair of giant planets near a 2:1 mean motion resonance \citep{2010ApJ...709..168A,2013ApJS..208....2W,2015A&A...577A.103K}. We perform a model comparison between a single, eccentric giant planet model and two giant planets on circular orbits for each system following \citet{2018MNRAS.480.2846B}. Kepler-65 shows no support for a two-giant-planet model, strengthening our earlier finding of a single moderately eccentric giant planet in the system. However, the model comparisons between a single giant and a pair of giants in the Kepler-25 and Kepler-68 systems are inconclusive but suggestive of a pair of giants just outside 2:1 resonance. We therefore urge further RV monitoring of both systems to definitively rule out (or confirm) near-resonant giant planet pairs. 

Kepler-65 e ($e=0.28 \pm 0.07$) is one of the highest eccentricity giant planets discovered to date exterior to a system of compact SEASNs. The relatively high eccentricities of Kepler-65 e and Kepler-68 d ($e=0.11\pm0.03$) compared to the low ($e<0.1$) eccentricities found in most multiplanet systems suggest that the processes which generate moderate giant planet eccentricity are not necessarily barriers to maintaining coplanar multi-planet systems at short orbital periods. In fact, these relatively high eccentricities may help reconcile the broad distribution of eccentricities seen in RV-detected giant planets with the low eccentricities seen in transiting compact multis (Fig.~\ref{fig:wcje}). However, they still do not reach the very high eccentricities (0.7--0.9) of some long-period planets detected with RVs, as such orbits may directly destabilize inner planets.

\cite{2015ApJ...808...14M} point out that giant planets with P$\lesssim$1 year and moderate eccentricities can be produced by high eccentricity migration channels; however, this evolutionary pathway destroys any compact interior planetary system. On the other hand, \cite{2008ApJ...686..580C} show that giant planets processed via scattering yields a broad range of eccentricities, including producing moderate eccentricity giant planets down to semi-major axes of a few tenths of an AU. We thus tentatively suggest that these systems experienced giant planet scattering events that did not disrupt the dynamically well-separated interior planets after their formation, rather than high eccentricity migration. 

Since we expect giant planets to form rapidly to allow sufficient time for runaway gas accretion before the protoplanetary disk dissipates \citep{1987Icar...69..249L,1996Icar..124...62P,2009Icar..199..338L}, we note that pebble accretion theories of SEASN formation \citep[e.g.,][]{2017AREPS..45..359J,2017ASSL..445..197O} could be disfavored in systems such as these with exterior giant planets on a wide range of radii that may halt pebble inflow \citep{1986ApJ...309..846L,2014A&A...572A..35L,2016Icar..267..368M}. However, the influence on giant planets in pebble accretion theories remains uncertain \citep[c.f.,][]{2011MNRAS.417.1236H,2012A&A...546A..18M}. The discovery of giant planets exterior to compact multiplanet systems may also disfavor large-scale migration of the inner bodies, despite the near-resonance observed in Kepler-25 \citep{2015ApJ...800L..22I}.

Overall, measuring the prevalence and eccentricities of giant planets exterior to compact SEASN systems puts constraints on the formation channels of the inner planets. 
Additional long-term monitoring of SEASN systems is needed to increase the number of detected outer companions and the number of systems with strong limits on the non-existence of giant planets. An increase in our understanding of the statistical distribution of exterior giants will provide further insight into both the giant planets' own dynamical histories and their influence on compact SEASN systems.

\acknowledgements

We thank the Kepler and Gaia teams for years of work making these precious datasets possible.
This research has made use of NASA's Astrophysics Data System, the Exoplanet Orbit Database, and the Exoplanet Data Explorer at exoplanets.org. M.R.K acknowledges support from the NSF Graduate Research Fellowship, grant No. DGE 1339067. L.M.W. acknowledges support from the Beatrice Watson Parrent Fellowship, the Trottier Family Foundation, and the Levy family.
The authors wish to recognize and acknowledge the very significant cultural role and reverence that the summit of Maunakea has long had within the indigenous Hawaiian community. We are most fortunate to have the opportunity to conduct observations from this mountain.

\facilities{\emph{Keck} (HIRES), \emph{Kepler}}

\software{\texttt{radvel} \citep{2018PASP..130d4504F}}


\appendix

\begin{longdeluxetable}{rrrr}
\tablecaption{Kepler-25 HIRES RVs\label{table:k25rvs}}
\tablehead{  \colhead{Time } & \colhead{RV } & \colhead{$\sigma$} & \colhead{S-value} \\
 \colhead{ [BJD-2454900]} & \colhead{m s$^{-1}$ } & \colhead{m s$^{-1}$} & }
{\def\arraystretch{0.8} \startdata
467.103&-7.25&3.84 &     0.133      \\
476.963&3.64&3.27 &      0.132     \\
477.950&-2.84&3.17 &      0.134     \\
533.942&14.53&3.65 &      0.127     \\
796.949&13.33&3.42 &      0.131     \\
797.952&2.34&3.67 &      0.130     \\
798.949&-4.09&3.46 &     0.130      \\
800.028&10.93&3.55 &      0.131     \\
834.053&-0.47&3.06 &      0.132     \\
834.942&-2.08&2.92 &     0.131      \\
835.987&-13.09&3.24 &     0.132      \\
839.050&-2.54&4.08 &     0.132      \\
851.936&-15.48&3.34 &      0.133     \\
852.794&-18.31&2.96 &    0.132       \\
859.966&6.28&2.94 &      0.133     \\
860.810&-4.88&3.10 &      0.133     \\
861.103&-3.76&2.98 &     0.133      \\
861.775&-0.48&3.61 &     0.130      \\
862.110&2.11&3.35 &       0.131    \\
862.891&-2.22&2.56 &     0.132      \\
863.841&-6.79&2.96 &    0.130       \\
868.853&-8.20&3.26 &     0.096      \\
869.944&9.28&3.54 &     0.124      \\
882.065&-5.56&4.07 &     0.129      \\
887.763&-0.92&3.71 &    0.130       \\
888.938&-4.86&3.29 &      0.128     \\
889.798&1.53&3.60 &      0.128     \\
889.805&4.00&4.06 &      0.129     \\
890.756&-2.39&3.27 &     0.129     \\
891.933&-0.20&3.31 &     0.128     \\
892.767&3.39&3.15 &       0.128   \\
894.926&17.73&3.71 &       0.126   \\
896.893&16.41&3.38 &        0.128  \\
898.759&4.48&3.38 &        0.127  \\
906.789&15.13&3.49 &     0.126       \\
908.797&20.58&3.12 &      0.127      \\
910.735&7.32&3.49 &      0.127      \\
914.927&15.07&3.57 &     0.126       \\
1003.699&10.62&3.48 &     0.127       \\
1179.873&-0.50&3.93 &   0.122         \\
1180.110&4.38&3.55 &      0.131      \\
1214.852&4.21&3.84 &      0.135      \\
1234.930&-11.01&3.27 &     0.130       \\
1239.028&-2.79&3.54 &      0.131      \\
1245.944&-1.82&4.20 &    0.131        \\
1247.807&-5.27&3.46 &    0.130        \\
1248.840&-6.27&3.87 &      0.129      \\
1252.963&15.96&3.70 &       0.128     \\
1263.794&23.03&3.54 &    0.129        \\
1294.822&7.34&4.06 &      0.127      \\
1575.858&-12.85&3.69 &     0.133       \\
1619.908&17.03&3.91 &     0.129       \\
1946.047&-1.94&3.33 &    0.131       \\
1946.900&-21.44&3.50 &      0.132     \\
1966.904&-10.92&3.84 &      0.122     \\
1984.002&3.68&3.92 &     0.127      \\
2010.879&1.79&3.91 &     0.127      \\
2280.055&14.22&3.69 &      0.128     \\
2301.960&-11.52&3.73 &     0.133      \\
2303.050&-20.87&3.61 &       0.132    \\
2308.083&-4.48&3.87 &    0.125       \\
2311.917&4.81&3.68 &      0.134     \\
2318.056&-5.55&3.62 &     0.128      \\
2329.923&6.56&3.85 &     0.133      \\
2332.886&-0.78&4.21 &    0.131       \\
2342.971&-9.59&3.98 &    0.130       \\
2345.851&2.25&4.09 &     0.131      \\
2696.075&-10.85&3.85 &    0.128       \\
2778.791&1.93&4.21 &     0.117      \\
3046.032&5.98&3.44 &     0.133      \\
3427.758     &   2.79     &  3.76 & 0.135\\
\enddata}
\end{longdeluxetable}

\begin{longdeluxetable}{rrrr}
\tablecaption{Kepler-65 HIRES RVs\label{table:k65rvs}}
\tablehead{  \colhead{Time } & \colhead{RV } & \colhead{$\sigma$} & \colhead{S-value} \\
 \colhead{ [BJD-2454900]} & \colhead{m s$^{-1}$ } & \colhead{m s$^{-1}$} & }{\def\arraystretch{0.8}
\startdata
 797.9785 & -1.46 & 3.03	  &     0.1330      \\
 798.9780 & 0.13 & 2.76	  &     0.1340      \\
 800.0428 & -5.06 & 2.92	  &     0.1350      \\
 835.9620 & -6.07 & 2.64	  &     0.1300      \\
 839.0158 & -13.93 & 3.09	  &   0.1330        \\
1950.0654 & -13.73 & 3.60	  &    0.1333      \\
1962.0654 & -7.68 & 3.90	  &   0.1340       \\
1965.0787 & -5.94 & 3.70	  &   0.1372       \\
1965.7623 & -7.32 & 3.69	  &   0.1329       \\
1966.9672 & -4.24 & 3.75	  &    0.1306      \\
1972.9521 & 5.86 & 3.60  &      0.1290    \\
1980.9310 & -4.06 & 3.12	  &   0.1360       \\
1991.8251 & 13.12 & 3.14	  &   0.1361       \\
1993.0213 & 1.71 & 3.20	  &     0.1349     \\
1994.9791 & 5.55 & 3.20	  &      0.1354    \\
2006.9404 & 12.55 & 3.57	  &  0.1350        \\
2010.8677 & 4.34 & 3.69	  &     0.1375     \\
2012.9413 & 8.11 & 3.59	  &      0.1290    \\
2101.7114 & 1.92 & 3.72	  &     0.1305     \\
2251.0331 & 18.43 & 3.40	  &  0.1338        \\
2279.9614 & 18.66 & 3.43	  &   0.1288       \\
2298.0195 & 25.93 & 3.20	  &   0.1312       \\
2303.0104 & 31.70 & 3.55	  &      0.1331    \\
2311.0792 & 20.51 & 3.57	  &   0.1344       \\
2316.0349 & 15.53 & 3.36	  &    0.1331      \\
2322.0992 & 11.10 & 3.44	  &    0.1356      \\
2323.1172 & 22.54 & 3.18	  &   0.1339       \\
2329.0174 & 9.84 & 3.37	  &    0.1350      \\
2329.9147 & 5.22 & 3.31	  &     0.1360     \\
2330.7927 & 19.60 & 3.41	  &     0.1369     \\
2332.0814 & 16.24 & 3.29	  &   0.1355       \\
2332.8916 & 12.18 & 3.53	  &   0.1355       \\
2334.1023 & 8.54 & 3.31	  &     0.1323     \\
2335.9536 & 2.34 & 3.20	  &    0.1357      \\
2336.8389 & 14.48 & 3.22	  &   0.1367       \\
2340.0231 & 8.16 & 2.92	  &    0.1361      \\
2344.8161 & 8.01 & 3.12	  &     0.1383     \\
2345.9187 & -9.75 & 3.04	  &   0.1380       \\
2347.0604 & -14.91 & 3.38	  &   0.1359       \\
2353.9886 & 5.19 & 3.01	  &    0.1373      \\
2354.9137 & -6.50 & 3.17	  &     0.1351     \\
2356.0186 & -1.05 & 3.48	  &   0.1188       \\
2362.8601 & -22.99 & 3.38	  &  0.1377        \\
2364.9807 & -9.91 & 3.21	  &  0.1378        \\
2385.7684 & -19.79 & 3.40	  &  0.1335        \\
2390.8331 & -9.75 & 3.41	  &  0.1337        \\
2390.9402 & -5.77 & 3.85	  &   0.1226       \\
2394.8182 & -5.33 & 3.04	  &   0.1364       \\
2396.9131 & -22.19 & 3.32	  &  0.1321        \\
2398.7801 & -13.78 & 3.18	  &  0.1334        \\
2426.7046 & -4.02 & 3.38	  &  0.1330        \\
2426.8396 & -10.35 & 3.63	  &  0.1340        \\
2453.6800 & -6.67 & 3.15	  &   0.1374       \\
2453.7663 & -17.74 & 3.56	  &   0.1347       \\
2454.6923 & -0.05 & 3.18	  &   0.1584       \\
2454.7650 & -7.60 & 3.49	  &     0.1357     \\
2455.6847 & -7.21 & 3.45	  &   0.1352       \\
2455.7584 & -7.93 & 3.29	  &   0.1346       \\
2478.6966 & -10.71 & 3.75	  &  0.1214        \\
2540.1678 & 21.70 & 3.28	  &   0.1312       \\
2578.1258 & 16.72 & 3.70	  &  0.1340        \\
2620.9898 & 9.11 & 3.28	  &      0.1350    \\
2662.0773 & -7.37 & 3.14	  &    0.1324      \\
2669.9411 & -17.18 & 4.17	  &    0.1335      \\
2680.0415 & -3.02 & 3.59	  &    0.1336      \\
2684.0287 & -26.03 & 3.77	  &    0.1326      \\
2697.0877 & -19.14 & 4.43	  &    0.1359      \\
2772.8783 & 11.25 & 3.17	  &    0.1352      \\
2988.0677 & -7.14 & 3.56	  &    0.1332      \\
3027.0476 & 2.71 & 3.37	  &      0.1337    \\
\enddata}
\end{longdeluxetable}

\begin{longdeluxetable}{rrrr}
\tablecaption{Kepler-68 HIRES RVs\label{table:k68rvs}}
\tablehead{  \colhead{Time } & \colhead{RV } & \colhead{$\sigma$} & \colhead{S-value} \\
 \colhead{ [BJD-2454900]} & \colhead{m s$^{-1}$} & \colhead{m s$^{-1}$} & }
{\def\arraystretch{0.8}
\startdata
413.082 &  -6.16 & 1.47 &   0.140        \\
419.109 &  -9.77 & 1.72 &    0.135       \\
422.051 & -12.09 & 1.26 &     0.138      \\
472.983 & -11.05 & 1.18 &    0.138       \\
477.929 &  -9.73 & 1.21 &    0.139       \\
481.000 & -17.54 & 1.27 &     0.138      \\
496.963 & -14.63 & 1.52 &    0.140       \\
512.923 & -18.64 & 1.19 &    0.137       \\
526.913 & -16.56 & 1.20 &    0.143       \\
531.784 & -10.77 & 1.19 &     0.140      \\
534.870 & -23.21 & 1.21 &      0.144     \\
534.876 & -24.48 & 1.20 &     0.144      \\
535.931 & -19.72 & 1.33 &     0.138      \\
536.968 & -15.75 & 1.69 &     0.135      \\
536.975 & -21.13 & 1.76 &     0.131      \\
537.940 & -15.08 & 1.32 &      0.143     \\
537.950 & -14.53 & 1.28 &      0.145     \\
538.991 & -15.84 & 1.27 &     0.139      \\
539.003 & -16.08 & 1.39 &      0.141     \\
539.924 & -19.47 & 1.19 &     0.144      \\
539.932 & -21.57 & 1.31 &      0.142     \\
540.971 & -18.38 & 1.21 &     0.143      \\
540.980 & -21.99 & 1.24 &     0.141      \\
555.810 & -22.84 & 1.28 &   0.148        \\
590.830 & -18.38 & 1.36 &     0.147      \\
772.026 & 17.17 & 1.12 &       0.138    \\
772.998 & 14.20 & 1.29 &      0.137     \\
773.996 & 18.22 & 1.37 &      0.139      \\
796.974 & 21.13 & 1.30 &     0.140       \\
797.964 & 23.13 & 1.43 &     0.140       \\
798.962 & 16.30 & 1.28 &      0.139      \\
822.995 & 21.99 & 1.40 &     0.137       \\
824.034 & 27.13 & 1.47 &     0.137       \\
828.901 & 22.75 & 1.51 &     0.125       \\
834.064 & 24.72 & 1.35 &     0.137       \\
834.951 & 25.41 & 1.31 &     0.138       \\
835.975 & 24.25 & 1.32 &      0.138      \\
839.034 & 23.77 & 1.41 &     0.137       \\
851.797 & 18.56 & 1.44 &    0.138        \\
852.105 & 17.53 & 1.33 &    0.134        \\
852.779 & 15.36 & 1.36 &     0.138       \\
859.975 & 18.30 & 1.32 &    0.139        \\
861.076 & 21.53 & 1.18 &     0.139       \\
861.842 & 15.24 & 1.23 &     0.134       \\
863.033 & 18.81 & 1.29 &     0.132       \\
863.851 & 14.05 & 1.26 &     0.139       \\
882.908 & 15.18 & 1.35 &     0.137       \\
895.024 &  7.12 & 1.53 &     0.139       \\
914.736 & 12.55 & 1.19 &    0.146        \\
1177.045 & -16.44 & 1.34 &   0.136         \\
1198.094 & -12.64 & 1.44 &   0.137         \\
1198.829 & -13.56 & 1.42 &   0.140         \\
1202.008 &  -6.61 & 1.30 &  0.138          \\
1214.872 & -15.53 & 1.30 &    0.139        \\
1245.875 & -10.22 & 1.39 &   0.141         \\
1248.929 &  -4.14 & 1.17 &   0.144         \\
1251.061 & -11.05 & 1.31 &    0.142        \\
1253.983 &  -9.76 & 1.21 &   0.146         \\
1274.827 &  -1.96 & 1.40 &  0.147          \\
1445.155 &  17.07 & 1.36 &   0.138         \\
1575.844 &   9.71 & 1.42 &  0.139          \\
1598.070 &   2.62 & 1.36 & 0.138           \\
1613.044 &  -2.67 & 1.36 &  0.141          \\
1619.920 &  -0.83 & 1.27 & 0.143           \\
1946.882 &  10.49 & 1.36 &   0.137        \\
1992.760 &   7.75 & 1.28 &  0.146         \\
1992.766 &   8.52 & 1.25 &   0.145        \\
1992.772 &  10.66 & 1.26 &   0.145        \\
2006.735 &  11.57 & 1.24 &  0.145         \\
2006.741 &  14.37 & 1.31 &  0.145         \\
2006.747 &  14.08 & 1.27 &    0.144       \\
2303.075 & -2.44 & 1.27 &     0.136      \\
2303.083 & -2.39 & 1.27 &    0.134       \\
2318.103 & -2.05 & 1.36 &    0.140       \\
2333.059 & -15.79 & 1.38 &    0.138       \\
2333.066 & -10.99 & 1.44 &   0.139        \\
2721.887 &  14.31 & 1.21 &   0.146        \\
2721.895 &   8.94 & 1.27 &  0.146         \\
2751.902 &  12.20 & 1.40 &   0.146        \\
3045.018 & -22.38 & 1.30 &    0.140       \\
3102.985 & -19.95 & 1.39 &   0.144        \\
3427.765 & 5.31 & 1.35   &  0.143          \\
\enddata}
\end{longdeluxetable}

\begin{figure*}
\centerline{
\includegraphics[scale=0.4]{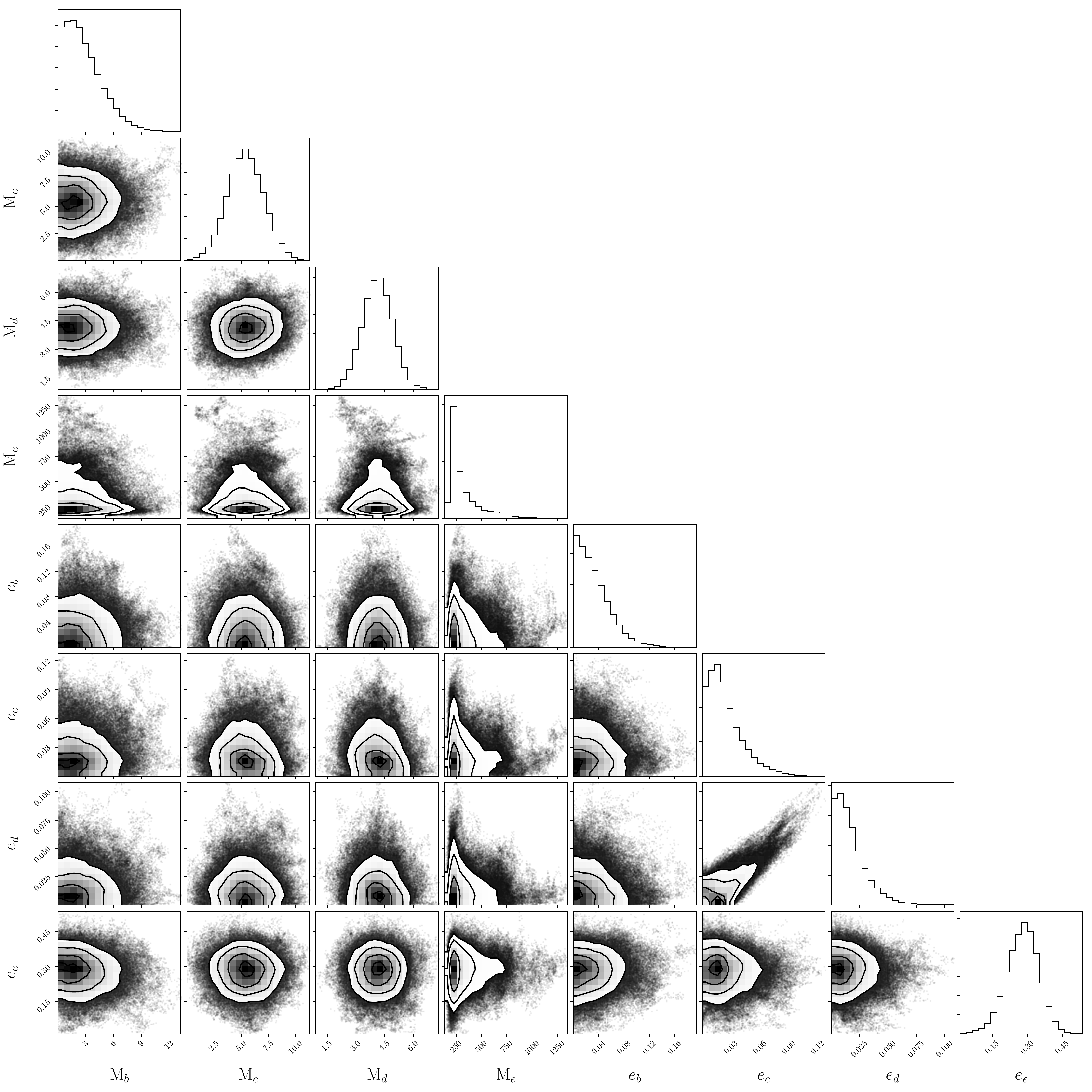}
}
\caption{Mass and derived eccentricity posteriors of Kepler-65's planets from a joint RV-\Kepler photometry fit. A significant TTV detection by the photodynamic model, combined with the low eccentricity prior results in mass constraints weakly degenerate with eccentricity.\label{fig:65corner}
}
\end{figure*}

\begin{figure*}
\centerline{
\includegraphics[scale=0.4]{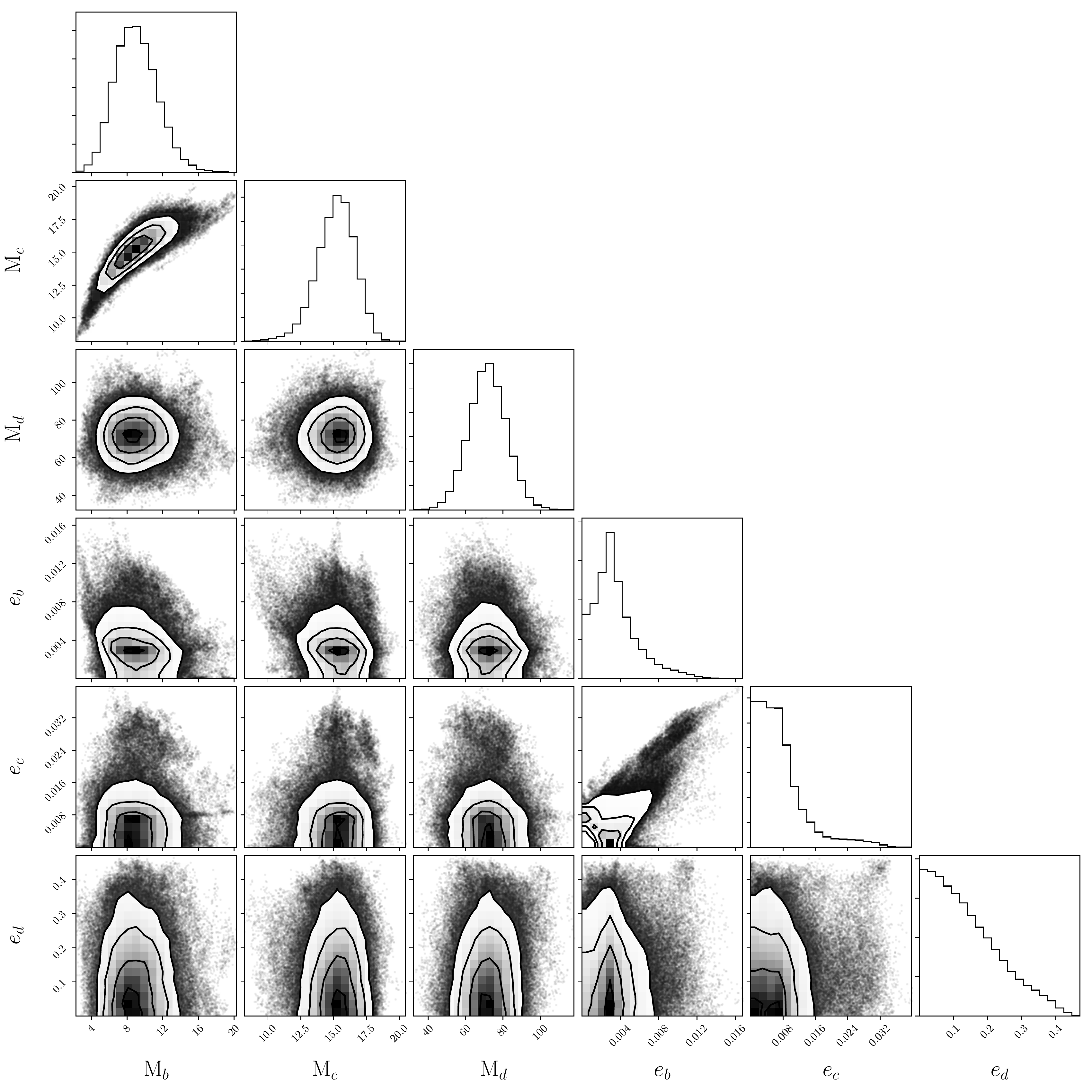}
}
\caption{Mass and derived eccentricity posteriors from a joint RV-\Kepler photometry fit of Kepler-25. A significant TTV detection by the photodynamic model, combined with the low eccentricity prior results in mass constraints weakly degenerate with eccentricity.\label{fig:25corner}
}
\end{figure*}

\end{document}